\documentclass{aa}
\usepackage[varg]{txfonts}
\usepackage{graphicx}
\usepackage{txfonts}
\usepackage{natbib}
\usepackage{verbatim}
\usepackage[pdftex]{color}
\usepackage{array}
\usepackage{ulem}

\bibpunct{(}{)}{;}{a}{}{,} % to follow the A&A style
\def\msun{M$_\odot$}
\def\lsun{L$_\odot$}

\usepackage{booktabs}
\newcommand{\iso}[1]{$^{#1}$}
\definecolor{ultramarine}{rgb}{0.07, 0.1, 0.6} 
\definecolor{myblue}{rgb}{0.07, 0.2, 0.6} 
\definecolor{dopal}{rgb}{.70, .25, .05}

\newcommand{\umco}[1]{\textcolor{dopal}{\bf{#1}}}
\newcommand{\one}[1]{\textcolor{myblue}{\bf{#1}}}
\begin{document}
\title{The formation of ultra-massive carbon-oxygen core white dwarfs and their evolutionary and pulsational properties}
%resulting from
%  single stellar evolution}

\author{Leandro G. Althaus\inst{1,2}, 
Pilar Gil Pons\inst{3},   
 Alejandro H. C\'orsico\inst{1,2},
 Marcelo Miller Bertolami\inst{1,2},
  Francisco De Ger\'onimo\inst{1,2},
  Mar\'ia E. Camisassa\inst{1,2,3}, 
        Santiago Torres\inst{3,4},
        Jordi Gutierrez\inst{3},
        Alberto Rebassa-Mansergas\inst{3,4}
        }
\institute{Grupo de Evoluci\'on Estelar y Pulsaciones. 
           Facultad de Ciencias Astron\'omicas y Geof\'{\i}sicas, 
           Universidad Nacional de La Plata, 
           Paseo del Bosque s/n, 1900 
           La Plata, 
           Argentina
           \and
           CCT - CONICET
           \and
           Departament de F\'\i sica, 
           Universitat Polit\`ecnica de Catalunya, 
           c/Esteve Terrades 5, 
           08860 Castelldefels, 
           Spain
           \and
           Institute for Space Studies of Catalonia, 
           c/Gran Capita 2--4, 
           Edif. Nexus 104, 
           08034 Barcelona, 
           Spain
           }
\date{Received ; accepted }

%Le saque las abreviaturas del abstract porque no estan permitidas (Ale- 22/6/2020)
%
\abstract{The existence of  ultra-massive white dwarf stars, $M_{\rm WD}  \gtrsim 1.05  M_\sun $,  has been reported 
in several studies. These white dwarfs are relevant for the role they play in type Ia Supernova explosions, the occurrence  of physical processes in the  asymptotic 
giant-branch phase, the  existence of high-field magnetic white dwarfs,  and the occurrence of double white dwarf mergers.}
{We explore the formation of ultra-massive, carbon-oxygen core white dwarfs  resulting from  
single  stellar evolution. We also study  their evolutionary and pulsational properties and compare them with those
of the  ultra-massive white dwarfs with oxygen-neon cores resulting from carbon burning in single progenitor stars, and with binary merger predictions. The aim is to provide a theoretical basis that can eventually help
to discern the core composition of ultra-massive white dwarfs and the scenario of their formation.}
{We   consider  two single-star evolution scenarios for the formation of ultra-massive carbon-oxygen core white dwarfs 
 that involve rotation of  the degenerate core after core helium burning and  reduced  mass-loss
 rates in massive asymptotic giant-branch stars.  We find that reducing standard mass-loss rates
 by a factor larger than $5-20$  yields the formation of carbon-oxygen cores  more massive than  $1.05\,M_\sun $   as a result of  the slow growth  of carbon-oxygen core mass during  the thermal pulses. We also performed a  series of evolutionary tests of solar-metallicity  models  with initial masses  between  4  and 9.5$\,M_{\odot}$, and with different core rotation rates. We 
find that ultra-massive carbon-oxygen core white dwarfs are formed even for the lowest rotation rates we analyzed, and that the
range of initial masses leading to these  white dwarfs widens as the rotation rate of the core
increases, whereas the initial mass range for the formation of oxygen-neon core white dwarfs decreases significantly. Finally, we compare our findings with  the predictions from  ultra-massive  white dwarfs resulting from the merger of two equal-mass  carbon-oxygen core white dwarfs, by  assuming complete mixing between them and a carbon-oxygen core for the merged remnant.}
{These two single evolution scenarios produce ultra-massive  white dwarfs with different carbon-oxygen profiles and different helium  contents, thus leading to distinctive signatures  in the period spectrum and 
mode-trapping properties of pulsating hydrogen-rich white dwarfs. The resulting
ultra-massive carbon-oxygen core white dwarfs 
evolve markedly slower than their oxygen-neon counterparts. }
{Our study strongly suggests the formation of ultra-massive white dwarfs with carbon-oxygen core from single stellar evolution. We find that both the evolutionary and pulsation properties
of these white dwarfs are markedly different from those of their oxygen-neon core counterparts and from those white dwarfs with carbon-oxygen core that
might result from double degenerate mergers. This can eventually be used to discern the core composition of ultra-massive white dwarfs and their formation scenario.}

\keywords{stars:  evolution  ---  stars: interiors  ---  stars:  white
  dwarfs --- stars: pulsations}
\titlerunning{Ultra-massive carbon-oxygen core white dwarfs}
\authorrunning{Althaus et al.}

\maketitle

\section{Introduction}
\label{introduction}

The white dwarf  (WD) state constitutes the final fate  for all of the
single  low-  and  intermediate-mass  stars.   WD  stars,  earth-sized
electron-degenerate stellar configurations, play  a major role for our
understanding of the formation and  evolution of stars, the history of
our Galaxy  and stellar  populations, and  the evolution  of planetary
systems                                                \citep[see][for
  reviews]{2008PASP..120.1043F,2008ARA&A..46..157W,
  2010A&ARv..18..471A,2019A&ARv..27....7C}.  Of particular interest is
the mass distribution of WDs.  Although there exists an intense debate
about its exact  distribution, it is generally agreed that  it shows a
main peak at $M_{ \rm WD} \sim 0.6 M_\sun $, and a smaller peak at the
tail  of  the distribution  around  $M_{\rm  WD}  \sim 0.82  M_\sun  $
\citep[e.g.][]{2013ApJS..204....5K,Jimenez2018}.   In   addition,  the
existence of ultra-massive WDs ($M_{\rm WD} \gtrsim 1.05 M_\sun $) has
been           reported          in           several          studies
\citep{2010MNRAS.405.2561C,2013MNRAS.430...50C,2013ApJ...771L...2H,
  2016MNRAS.455.3413K,2017MNRAS.468..239C,Hollands2020}.         Also,
\cite{2015MNRAS.452.1637R} report  the existence of a  clear excess in
the number  of hydrogen (H)-rich WDs  with masses near $1\,  M_\sun$.  The
interest in  ultra-massive WDs is motivated  by the role they  play in
type  Ia  Supernova,  the  occurrence of  physical  processes  in  the
Asymptotic Giant Branch  (AGB) phase, the  existence of high-field
magnetic  WDs, and the double  WD
mergers \citep{2015ASPC..493..547D,2020arXiv200614688R}.  In addition,
because of their very high  central densities, ultra-massive WDs offer
a test bed  to infer and explore the theory  of crystallization thanks
to the  capability of asteroseismology  to potentially probe  the very
deep   interior    of   these    stars   \citep[see][for    a   recent
  review]{2019A&ARv..27....7C}.

Ultra-massive  WDs are  thought to  be the  outcome of  single stellar
evolution of  progenitor stars with  masses higher than  6--9 $M_\sun$
that lose  their envelope  through winds before  the core  reaches the
Chandrasekhar mass (if this happens, the result is an electron-capture
supernovae, ECSNe). This initial mass threshold depends on metallicity
and  input physics  such  as the  treatment  of convective  boundaries
\citep[see][for   a    review]{2017PASA...34...56D}.    After   helium (He)
exhaustion, these  stars evolve to  the Super Asymptotic  Giant Branch
(SAGB), where core temperatures become high enough to start off-center
carbon ignition  under partially  degenerate conditions, leading  to a
carbon-burning  flash   with  associated   luminosity  up   to  10$^9$
\lsun.  This takes  place when  the carbon-oxygen  (CO) core  mass has
grown  to about  1.05$\,M_\sun$ and  before the  thermally-pulsing AGB
(TP-AGB)  phase  is   reached  (see,  e.g.  \citealt{garciaberro1994};
\citealt{2007A&A...476..893S,   2017PASA...34...56D}).   The   violent
carbon-ignition   phase  is   followed  by   the  development   of  an
inward-propagating convective flame that transforms the CO core into a
degenerate  oxygen-neon  (ONe)  core\footnote{In  some  cases,  carbon
  ignition results aborted and a CO  core surrounded by a ONe envelope
  could  be  expected  (\citealt{doherty2010},  \citealt{ventura2011},
  \citealt{denissenkov2013}).}
\citep{1997ApJ...485..765G,2005A&A...433.1037G,2006A&A...448..717S,2010MNRAS.401.1453D,2011MNRAS.410.2760V}.
%If stellar winds remove the H-rich envelope before the occurrence of electron captures, the ECSN outcome is avoided, and the star completes its thermally-pulsing SAGB phase to finally become a WD.
In this  scenario, ultra-massive WDs  with stellar masses  larger than
$M_{\rm  WD}  \gtrsim 1.05  M_\sun  $   composed of  $^{16}$O  and
$^{20}$Ne and traces  of $^{23}$Na  and $^{24}$Mg  are expected  to
emerge \citep{2007A&A...476..893S}.

An alternative scenario for the formation of ultra-massive WDs
has gained relevance in recent years. Indeed,
evidence has been mounting that a fraction of single ultra-massive WDs
could be the result of binary evolution channels.  Recent studies
point out toward a substantial  contribution of binary mergers to the
single WD  population \citep{2017A&A...602A..16T,2018MNRAS.476.2584M}.
In particular, \cite{2020A&A...636A..31T} conclude  that, on the basis
of binary population synthesis results,
%about 10--30$\%$ of all observable single WD are formed via binary mergers, and that 
about  30--45$\%$  of  all  observable single  WD  more  massive  than
0.9$\,M_\sun$ within 100 pc are  formed through binary mergers, mostly
via  the  merger of  two  WDs.  In  addition,  based on  the  velocity
distribution  of  high-mass  WDs   in  the  range  0.8--1.3$\,M_\sun$,
\cite{2020ApJ...891..160C}  estimate that  the  fraction of  double-WD
mergers in their sample amounts to  about 20 $\%$.
The result of such WD mergers remains a matter of debate. According to 
\cite{2007MNRAS.380..933Y} and \cite{2009A&A...500.1193L}, the
possibility that  the merger remnants avoid off-center carbon burning and become single ultra-massive WDs characterized by a CO core (hereinafter referred to as UMCO WDs) cannot 
be discarded. However, recent studies based on one-dimensional post merger evolutionary models \citep{2020arXiv201103546S} predict ultra-massive WDs with ONe cores as a result of off-center carbon burning in the merged remnant, see also \cite{2012ApJ...748...35S}.

% Yo pondria esto: 
%The outcome of such double-WD mergers  is the  formation of single  
%ultra-massive, CO-core (UMCO) WDs, \cite{2012ApJ...749...25G}. 
%{\bf In a recent study, \citep{2020arXiv201103546S} obtains 
%ultra-massive WDs with  ONe cores resulting from WD mergers.}

Fortunately, several ultra-massive H-rich  WDs (DA WDs) exhibit
$g$(gravity)-mode               pulsational              instabilities
\citep{2005A&A...432..219K,2010MNRAS.405.2561C,2013MNRAS.430...50C,
  2013ApJ...771L...2H,2017MNRAS.468..239C,2019MNRAS.486.4574R},     so
their  internal structure  could be  probed through  asteroseismology.
The study of the predicted  pulsational patterns of UMCO and ONe-core
WDs could eventually  help to distinguish between both  types of stars
and place  constrains to  the different channels  in the  formation of
these stars.  In this sense, based on new evolutionary models computed
by \cite{2019A&A...625A..87C}  for the evolution of  ONe-core WDs that
consider  new  phase  diagrams  for ONe  phase  separation  core  upon
crystallization,  \cite{2019A&A...621A.100D}   have  shown   that  the
features  found in  the period-spacing  diagrams  could be  used as  a
seismological  tool  to  discern  the core  composition  of  pulsating
ultra-massive   WDs.   Also,   on   the   basis   of   these   models,
\cite{2019A&A...632A.119C}    have     carried    out     the    first
asteroseismological analysis of ultra-massive  ZZ Ceti stars, focusing
in particular on the stars BPM~37093, GD~518, and SDSS~J0840+5222.

In this paper, we explore  two possible single evolution scenarios for
the formation  of UMCO WDs.  One
scenario  exploits wind  rate uncertainties  during the  TP-AGB phase,
see, e.g. \cite{decin2019}, and involves  the reduction of these rates
below  the values  given  by standard  prescriptions. The  alternative
scenario involves  the occurrence of  rotation in degenerate  CO cores
expected from  the contraction  of the  core at the  onset of  the AGB
phase  following   central  He   exhaustion,  \cite{dominguez1996}
(hereinafter Dom96). We will show that both  the evolutionary and pulsational 
properties of the UMCO WDs  formed through these two single  evolution scenarios are
markedly different from those of their ONe-core counterparts and from those white dwarfs with carbon-oxygen core that might result from double degenerate mergers.
This  can eventually  be  used  to discern  the  core composition  of
ultra-massive WDs and their origin.

The paper is  organized as follows.  In  Sect.~\ref{codes} we describe
the evolutionary  and pulsational  codes. In  Sect.~\ref{scenarios} we
present  details about  the scenarios  we  explored that  lead to  the
formation    of     UMCO    WDs.     In     Sect.~\ref{results}    and
~\ref{pulsation_results} we describe  the evolutionary and pulsational
properties,   respectively,  of   the  resulting   WDs.  Finally,   in
Sect.~\ref{conclusions} we summarize the main findings of the paper.

\section{Evolutionary and pulsational codes}
\label{codes}

Three different but complementary codes have been applied in the analysis done in this work. In what follows we summarize the main input physics used in them.

The structure  and composition  of the UMCO  WD progenitors  have been
computed   with  the   Monash-Mount   Stromlo  code   \textsc{monstar}
(\citealt{wood1987}, \citealt{frost1996};   \citealt{campbell2008}),   and  presented   in
\cite{gilpons2013,gilpons2018}.   Here  we  summarize the  main  input
physics used. The  mixing-length to pressure scale  height quotient is
$\alpha=1.75$.  Usually, convective  boundaries with  \textsc{monstar}
are determined  using the Schwarzschild  criterion and the  search for
neutrality  approach (\citealt{castellani1971},  \citealt{frost1996}).
As recent calculations with \textsc{monstar} \citep[see e.g.][and references therein]{doherty2017}, mass-loss  rates  during the  RGB  phase  follow the  prescription  of
\citet{reimers1975} with $\eta$=0.4. Mass loss during this part of the evolution is very modest ($\lesssim$ 0.05 \msun) and thus its effects on the overall evolution are minor. For the AGB phase,  we use the standard wind-rate
prescription  of  \cite{vassiliadis1993}.  Overshooting has  not  been
considered. 
Relativistic and ion gases are treated using the fitting formulae by \citet{beaudet1971}. Interior stellar opacities are from \citet{1996ApJ...464..943I}. Low-temperature opacity tables are from  \citet{lederer2009,marigo2009}, and take into account variable composition effects.

For  this work,  we  have  implemented a  simplified  approach to  the
effects of rotation following the  treatment described in Dom96, which
is  based  on  \citet{kippenhahn1970}.   This  treatment  provides  an
approximation  to   the  inherently  multidimensional   phenomenon  of
rotation  for one-dimensional  evolutionary codes.  In particular,  it
captures the  main effect of the  decrease of core pressure  caused by
the increase  of angular velocities  expected from the  compression of
the  CO  core  at  the  onset of  the  AGB  phase.  Specifically,  the
hydrostatic  equilibrium  equation  is  modified inside  the  core  by
introducing the rotation parameter $f$ as follows:

\begin{equation}
    \frac{dP}{dM_r}=-\frac{GM_r}{4\pi\:r^4}(1-f),
\end{equation}{}

\noindent where $f$ is related to the critical velocity, $\sqrt f=\omega/\omega_{\rm crit}$. 
For the sake of simplicity, and also following Dom96, $f$ is taken to be constant. 

 More  sophisticated   implementations  of   rotation  exist   in  the
 literature     (see,      for     instance,     \citealt{maeder2000};
 \citealt{heger2000}; \citealt{farmer2015}; \citealt{limongi2018}; and
 \citealt{paxton2019}).  However,  the  physics  of  internal  angular
 momentum transport in stars is not well understood, and sophisticated
 models that allow a proper assessment of the evolution of the angular
 momentum in evolved low-mass stars  fail to predict the observed core
 rotation   of    early   red    giants   and   clumps    stars,   see
 \cite{2014ApJ...788...93C}.   In fact,  all these  approaches neglect
 the  basic  dimensionality  of  a rotating  structure,  for  which  a
 two-dimensional approach  seems to be the  minimum accuracy requisite
 \citep{2013A&A...552A..35E}.  Hence,  we  rely instead  on  the  much
 simpler model we have described, which is enough for our purposes. We
 stress that we do not attempt here to follow the evolution of angular
 momentum  in prior  evolutionary stages,  but simply  to explore  the
 expectation for  the CO-core  masses and to  achieve initial  UMCO WD
 configurations for our evolving models by assuming different rotation
 rates during the AGB phase. As  shown in Dom96, this simple treatment
 to simulate the  presence of rotation captures  the expected decrease
 in maximun  temperature in  response to pressure  decrease, favoring
 the  formation of  UMCO cores.  More elaborated  stellar models  that
 include the  variation of  the angular momentum  content of  the star
 predict also a  significant increase in the CO-core mass  as a result
 of core rotation,  see Table 1 of  \cite{2013A&A...553A..24G} for the
 case of more  massive stars than studied here. Finally,  it should be
 mention  that when  studying SAGB  stars, \citet{farmer2015}  did not
 point to marked  differences in important aspects  of carbon burning,
 such as  ignition point  masses, or  flame quenching  locations, when
 rotation is  implemented. This  is probably  justified by  the strong
 influence  of  the  magnetic  torques  implemented  in  the  work  by
 \citet{farmer2015},  which  significantly  inhibited the  spin-up  of
 model cores.  We note that our  knowledge of this phenomenon  in stellar
 evolution is still  poorly known and further research has  to be done
 to  completely  understand  the  behavior  of  magnetic  fields  and
 rotation \citep{2019A&A...625A..89G}.

The initial composition of our models was taken from \citet{grevesse1996}, and \textsc{monsoon}, the postprocessing code developed at Monash University \citep{can93,lug04,doh14}, was used to perform detailed nucleosynthetic calculations. The specific version we used includes 77 species, up to \iso{32}S and Fe-peak elements. It also includes a 'g' particle \citep{lug04}, which is a proxy for s-process elements. The neutron-sink approach (\citet{jorissen1989,lugaro2003,herwig2003}) allows to account for eventual neutron captures on nuclides that are not present in our network. 
Most nuclear reaction rates are from the \textsc{JINA} reaction library \citep{cyb10}. p-captures for the NeNa-cycle and MgAl chain are from \citet{iliadis2001}, p-captures on \iso{22}Ne are from \cite{hale2002}, $\alpha$-captures on \iso{22}Ne are from \citet{karakas2006}, and p-captures o \iso{23}Na are from \cite{hale2004}. 

The WD evolutionary models used in  this work have been computed using
the {\tt LPCODE}  stellar evolutionary code that has  been widely used
to  study  the evolution  of  low-mass  and WD  stars  \citep[see][for
  details  about  the code]  {2003A&A...404..593A,2005A&A...435..631A,
  2015A&A...576A...9A,  2016A&A...588A..25M}.  {\tt  LPCODE} has  been
tested  and  calibrated  with  other  stellar  evolutionary  codes  in
different   evolutionary  stages,   such  as   the  red   giant  phase
\citep{2020A&A...635A.164S,  2020A&A...635A.165C}  and  the  WD  stage
\citep{2013A&A...555A..96S}.
%see       \cite{2008A&A...491..253M},      \cite{2010Natur.465..194G},
%\cite{2010ApJ...717..897A},                \cite{2010ApJ...717..183R},
%\cite{2011ApJ...743L..33M},                \cite{2011A&A...533A.139W},
%\cite{2012MNRAS.424.2792C},                \cite{2013A&A...555A..96S},
%\cite{2013A&A...557A..19A},       \cite{2016A&A...588A..25M},      and
%\cite{2017ApJ...839...11C} for  different applications  and subsequent updates.  
Relevant  for  the   present  work,  {\tt  LPCODE}   considers  a  new
full-implicit  treatment  of  time-dependent  element  diffusion  that
includes  thermal and  chemical diffusion  and gravitational  settling
\citep{2020A&A...633A..20A},  outer  boundary conditions  provided  by
nongray            model            atmospheres            \citep[for
  references]{2012A&A...546A.119R,2017ApJ...839...11C,2018MNRAS.473..457R},
and  a full  treatment of  energy  sources, in  particular the  energy
contribution ensuing  from phase  separation of core  chemical species
upon crystallization.   The treatment  of crystallization is  based on
the most  up-to-date phase diagrams of  \cite{2010PhRvL.104w1101H} for
dense  CO mixtures,  and  that of  \cite{2010PhRvE..81c6107M} for  ONe
mixtures.   In   this  work,   we   have   not  considered   $^{22}$Ne
sedimentation.  Recently, {\tt LPCODE} has also been used to calculate
a  grid of  ultra-massive ONe  WDs with  stellar masses  in the  range
$1.10-1.29\,M_{\odot}$  \citep{2019A&A...625A..87C} based  on detailed
chemical profiles  as given by $9-10.5\,M_{\odot}$  single progenitors
evolved  through the  semi-degenerate  carbon burning  and the  TP-AGB
phases \citep{2010A&A...512A..10S}.
%This provides  realistic and consistent chemical profiles for the resulting ultra-massive ONe WDs,
%including the  O/Ne inner profiles and the outer
%chemical stratification, in particular the mass of the He inter-shell
%built up during the SAGB,  a key issue as far as  the assessment of
%cooling times of these WDs at low luminosities is concerned. 
These  new ultra-massive  ONe models  include for  the first  time the
release of  energy and the  core chemical redistribution  ensuing from
the phase  separation of $^{16}$O and  $^{20}$Ne upon crystallization,
thus  substantially improving  previous  attempts  at modeling  these
stars. We want to mention that  during the WD regime, rotation has not
been considered.

\begin{figure*}
\centering
\includegraphics[width=1.0\linewidth]{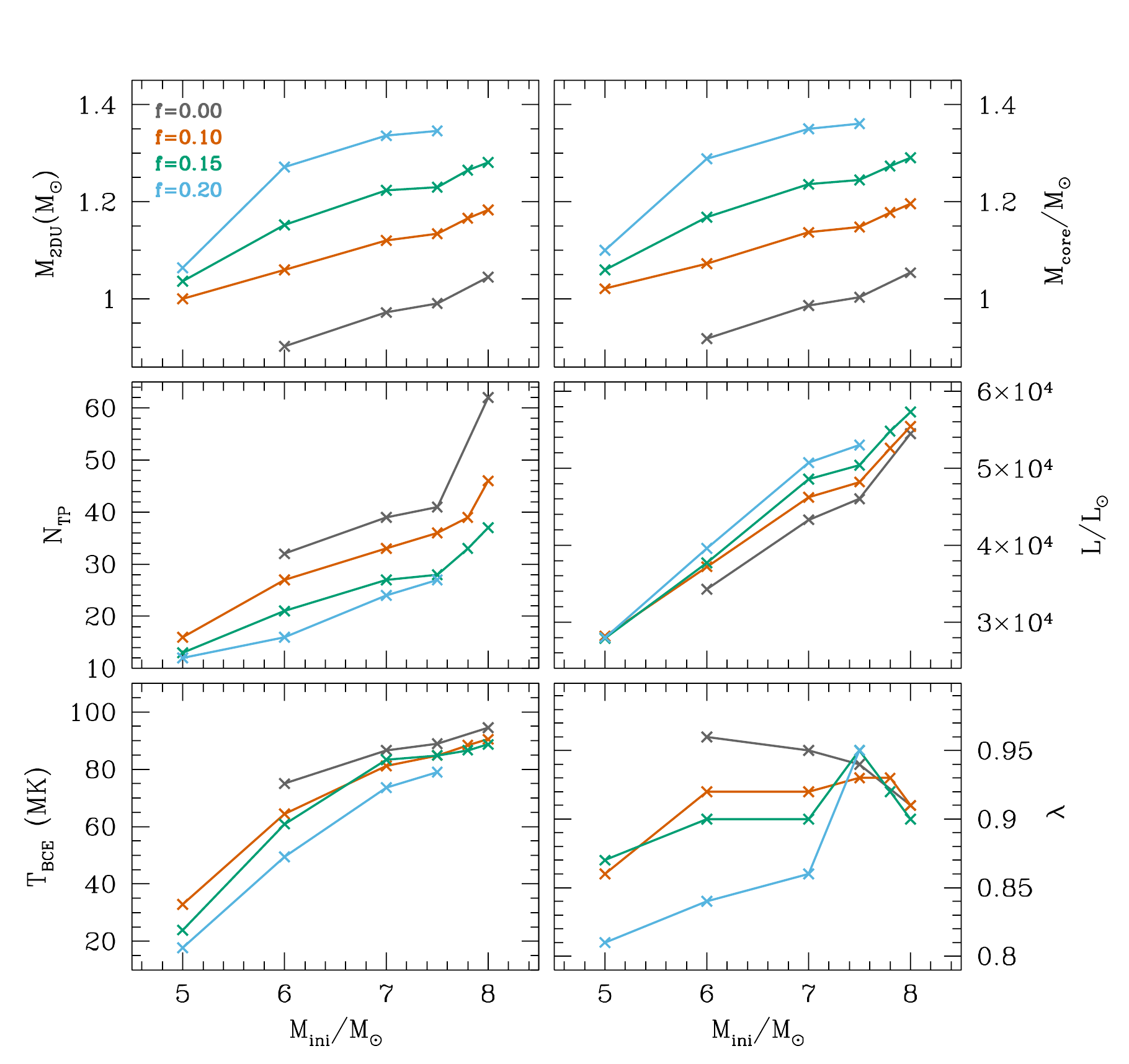}
\caption{Upper panels: H-exhausted core mass at the maximum advance of the SDU (left) and at the end of our calculations (right). Middle panels: number of thermal pulses of our sequences (left) and maximum surface luminosity values reached during the TP-AGB (right). Lower panels: Maximum temperature at the base of the convective envelope (left) and maximum TDU parameter (right). All values are given with respect to the initial model masses. Black lines and symbols represent cases with no rotation, orange, green and blue  represent, respectively, cases with rotation and $f$=0.1, 0.15, and 0.2.} 
        \label{fig:panelevo}
\end{figure*}

\begin{table*}[t]
    \centering
    \begin{tabular}{lcccccccccccccc}
%\begin{tabular}{p{0.2\textwidth}>{\centering}p{0.2\textwidth}>{\centering}p{0.2\textwidth}>{\centering}p{0.2\textwidth}>{\centering}p{0.2\textwidth}>{\centering}p{0.2\textwidth}>{\centering}p{0.2\textwidth}>{\centering}p{0.2\textwidth}>{\centering}p{0.2\textwidth}>{\centering}p{0.2\textwidth}>{\centering}p{0.2\textwidth}>{\centering}p{0.2\textwidth}>{\centering}p{0.2\textwidth}>{\centering}p{0.2\textwidth}}
    \hline
    \hline
    $M_\mathrm{ini}$ & TP-AGB & $f$ & $M_{\rm c,CHeB}$ & $M_{\rm c,SDU}$ & $M_{\rm c,f}$ & $M_{\rm env,f}$  & $N_{\rm TP}$ & $\tau_{\rm TP-AGB}$ & $\Delta t_{\rm IP}$ &     $T_{\rm BCE}$  & $L_{\rm max}$ & $\lambda$ & $C/O$ \\[1pt]
    \msun & & & \msun & \msun & \msun & \msun &  & Myr & yr & MK &  $ 10^3$\lsun&  \\[1pt]
    \hline \\[-4pt]    
%    \multicolumn{12}{c}{{\bf VW93}}\\[1pt]
     7.8  & VW, SCN & 0 & 1.673 & 1.016 & 1.029 & 1.681 & 47 & 0.095 & 3016 & 91.5 & 49.6 & 0.97 & 0.34 \\  
     7.8  & VW/10, SCN & 0 & 1.673 & 1.016 & 1.066 & 2.303 & 406 & 0.779 & 2289 & 95.2 & 55.3 & 0.93 & 2.29 \\    
     7.8  & VW/20, SCN & 0 & 1.673 & 1.016 & 1.112 & 2.732 & 771 & 1.107 & 2283 &  95.7 & 57.0 & 0.92 & 2.08 \\
     7.8  & VW/50, SCN & 0 & 1.673 & 1.016 & 1.171 & 4.229 & 1357 & 1.431 & 2274 & 96.7 & 63.3 & 0.93 & 1.04 \\
    % 7.8  & VW/100, SCN & 0 & 1.673 & 1.016 & 1.081 & 6.597 & 447 & 0.786 & 1534& 96.1 & & 0.83 & 0.33 \\    
     \hline \\[-4pt]
     7.8  & VW, Schw & 0 & 1.709 & 1.025 & 1.049 & 2.352 & 50 & 0.086 & 1727 & 92.5 & 50.7 & 0.62 & 0.17 \\    
     7.8  & VW/2, Schw & 0 & 1.709 & 1.025 & 1.062 & 2.576 & 92 & 0.137 & 1553 & 94.2 & 52.1 & 0.63 & 0.22 \\ 
     7.8  & VW/5, Schw & 0 & 1.709 & 1.025 & 1.102 & 2.909 & 234 & 0.264 & 1528 & 96.0 & 55.1 & 0.63 &  0.33\\ 
       \hline \\[-4pt]
     5.0  & VW, SCN & 0.00 & 1.036 & 0.863 & 0.877 & 1.405 & 26 & 0.193 & 26222 & 57.6 & 26.8 & 0.94 & 0.85 \\        
     5.0  & VW, SCN & 0.15 & 1.036 & 1.036 & 1.050 & 1.441 & 13 & 0.107 & 10075 & 23.9  & 27.9 & 0.87 & 0.58 \\    
     5.0  & VW, SCN & 0.20 & 1.064 & 1.064 & 1.105  & 1.485 & 12 & 0.126 & 12692 & 17.8 & 28.0 & 0.81 & 0.56 \\
     6.0  & VW, SCN & 0.00 & 1.268 & 0.902 & 0.918 & 1.632 & 32 & 0.201 & 6581 & 75.0 & 34.3 & 0.96 & 0.31 \\
     6.0  & VW, SCN & 0.10 & 1.268 & 1.060 & 1.076 & 1.660 & 27 & 0.092 & 5264 & 66.4 & 37.2 & 0.92 & 0.55 \\    
     6.0  & VW, SCN & 0.15 & 1.268 & 1.152 & 1.168 & 1.804 & 21 & 0.074 & 4826 & 57.7 & 37.7 & 0.90 & 0.58 \\    
     6.0  & VW, SCN & 0.20 & 1.268 & 1.272 & 1.289 & 1.817 & 16 & 0.056 & 3984 &  49.5 & 39.6 & 0.84 & 0.50 \\
     7.0  & VW, SCN & 0.00 & 1.526 & 0.972 & 0.974 & 2.341 & 35 & 0.133 & 3579 & 87.0 & 43.6 & 0.95 & 0.30 \\
     7.0  & VW, SCN & 0.05 & 1.526 & 1.038 & 1.054 & 1.830 & 34 & 0.123 & 3414 & 84.5 & 44.6 & 0.91 & 0.33 \\
     7.0  & VW, SCN & 0.10 & 1.526 & 1.120  & 1.137 & 1.755 & 33 & 0.117 & 3336  & 77.4 & 48.0 & 0.91 & 0.35 \\    
     7.0  & VW, SCN & 0.15 & 1.526 & 1.224 & 1.236 & 2.115 & 27 & 0.065 & 3171 & 77.4 & 48.6 & 0.90 & 0.29\\    
     7.0  & VW, SCN & 0.20 & 1.526 & 1.336 & 1.350 & 2.012 & 22 & 0.035 & 4201 & 73.6 & 56.7 & 0.86 & 0.18 \\ 
     7.5  & VW, SCN & 0.00 & 1.591 & 1.055 & 1.071 & 1.849 & 38 & 0.102 & 3741 & 89.0 & 46.0 & 0.97 & 0.30 \\    
     7.5  & VW, SCN & 0.10 & 1.591 & 1.134 & 1.148 & 1.606 & 36 & 0.092 & 3739 & 84.9 & 48.3 & 0.95 & 0.37 \\    
     7.5  & VW, SCN & 0.15 & 1.591 & 1.230 & 1.245 & 1.551 & 33 & 0.083 & 3686 & 84.8 & 50.4 & 0.95 &  0.41 \\
%     7.5  & VW, SCN & 0.20 & 1.591 & 1.346 & 1.371 & 1.819 & 25 & 0.034 & 3450 & 79.0 & 53.0 & 0.94 & 0.35 \\ 
     7.8  & VW, SCN & 0.10 & 1.684 & 1.166 & 1.178 & 1.752 & 39 & 0.084  & 3066 & 88.6 & 52.6 & 0.93 & 0.13 \\    
     7.8  & VW, SCN & 0.15 & 1.684 & 1.265 & 1.274 & 1.803 & 33 & 0.073 & 3141 & 86.7 & 54.8 & 0.92 & 0.31 \\    
     8.0  & VW, SCN & 0.00 & 1.735  & 1.106 & 1.116 & 2.102 & 45 & 0.089 & 3473 & 94.6 & 54.5 & 0.91 & 0.28 \\ 
     8.0  & VW, SCN & 0.10 & 1.735  & 1.183 & 1.196 & 1.768 & 45 & 0.087 & 4306 & 90.5 & 55.4 & 0.91 & 0.35 \\    
     8.0  & VW, SCN & 0.15 & 1.735 & 1.281 & 1.291 & 1.800 & 35 & 0.074 & 2768 & 88.7 & 57.3 & 0.90 & 0.35 \\    

            \hline \\[-4pt]
       %\vspace{0.1cm}
%        \multicolumn{13}{c}{{\bf Blo95}}\\[1pt]
       \\
    \end{tabular}
    \caption{Main characteristics of the TP-(S)AGB of our  models leading to UMCO WD. For comparative purposes, we also added some selected nonrotating models or altered stellar winds. $M_\mathrm{ini}$ corresponds to the initial mass. TP-AGB describes the main input physics during this evolutionary stage: VW refers to mass-loss rates by \citet{vassiliadis1993}, SCN to the implementation of the search for convective neutrality approach to the determination of convective boundaries (see main text for details), and Schw to the implementation of the strict Schwarzschild criterion.  $f$ corresponds to the rotation parameter ($f$=0 means there is no rotation).  
    $M_{\rm c,CHEB}$ and $M_{\rm c,SDU}$ are, respectively, the H-exhausted core masses at the end of core He-burning, and at the deepest advance of the second dredge-up. 
    $M_{\rm c,f}$ and $M_{\rm env,f}$ are the final core mass and the H-rich envelope mass left at the end of our calculations. 
    $N_{\rm TP}$, $\tau_{\rm TP-(S)AGB}$ and $\Delta t_{\rm IP}$ are, respectively, the number of thermal pulses, the duration of the TP-(S)AGB (given from the first thermal pulse until the end of our computations), and the maximum interpulse period in each sequence. $T_{\rm BCE}$, $L_{\rm max}$ and $\lambda$ are, respectively, the maximum temperature at the base of the convective envelope, the maximum luminosity during the TP-(S)AGB, and the maximum TDU efficiency parameter. $C/O$ is the ratio of final surface number abundances of carbon to oxygen. 
%    Pilar: recuerda aclarar en main text que los casos 8 \msun con f=0.02, 0.05 dan CONes. Las incluí como UMCO porque la zona del core rica en Ne era sólo de 0.18 \msun. La evolución de estas enanas se deja para un posible trabajo futuro.
    }
    \label{tab:evol}
\end{table*}

Finally, for the pulsation analysis of  our WD models, we employed the
adiabatic  version of  the {\tt  LP-PUL} pulsation  code described  in
\citet{2006A&A...454..863C}.  This code has  been employed recently by
\cite{2019A&A...621A.100D} to  study the  pulsation properties  of the
ONe-core      ultra-massive      WD       models      computed      by
\cite{2019A&A...625A..87C},   and  by   \cite{2019A&A...632A.119C}  to
perform  the first  asteroseismological analyses  of ultra-massive  ZZ
Ceti  stars. To  account for  the  effects of  crystallization on  the
pulsation spectrum of  $g$-modes, we adopt the  "hard sphere" boundary
conditions, which assume  that the amplitude of  the eigenfunctions of
$g$-modes  is drastically  reduced  below  the solid/liquid  interface
because of  the nonshear modulus of  the solid, as compared  with the
amplitude in  the fluid region  \citep[see][]{1999ApJ...526..976M}. In
our code, the inner boundary condition  is not the stellar center, but
instead  the mesh-point  corresponding  to  the crystallization  front
moving                toward                the                surface
\citep[see][]{2004A&A...427..923C,2005A&A...429..277C,2019A&A...621A.100D,2019A&A...632A.119C}.
The    Brunt-V\"ais\"al\"a    frequency     is    computed    as    in
\cite{1990ApJS...72..335T}.  The  computation of  the Ledoux  term $B$
---a  crucial amount  involved  in  the Brunt-V\"ais\"al\"a  frequency
calculation---  includes  the  effects  of  having  multiple  chemical
species that vary in abundance.

\section{The formation of UMCO  WDs}
\label{scenarios}

We explore in this section the formation of  UMCO WDs resulting from single stellar
evolution by focusing on the role of rotation and mass loss in the evolution of the degenerate CO cores of evolved AGB stars. We also considered the formation 
of UMCO WDs via a double-WD merger.
%We concentrate on the evolutionary aspects that provide us with initial chemical structures for our evolving UMCO WD sequences.

\subsection{Stellar rotation scenario}
\label{rotation}

We have performed a series of evolutionary tests for solar metallicity
models with initial masses between 4 and 9.5 $M_{\odot}$ and values of
the  rotation  parameter $f$ (rotation has been considered after core He burning)
of  0.02,  0.05,  0.1, 0.15,  0.20,  and
0.25.    They    correspond,    respectively,   to    the    following
$\omega/\omega_{\rm crit}$  values: 0.14, 0.22, 0.32,  0.39, 0.45, and
0.50. In this range of values we maintain the stability criterion that
the  ratio  of kinetic  to  gravitational  energy remains  below  0.14
\citep{durisen1975,2012sse..book.....K}, as  well as those  related to
the   stability   of   differential   rotation  (see   Dom96   for   a
discussion).

We now describe the evolution of the early AGB (E-AGB) and TP-AGB of models leading to UMCO WDs. Given the simplicity of our implementation of rotation, and that a thorough analysis of its effects on intermediate-mass models is beyond the scope of this work, our description merely aims to highlight the main properties of our sequences, and compare them to the results presented in Dom96.  
The main properties of rotating models leading to UMCO WD and, as a reference, some nonrotating ones of analogous masses, are summarized in Table \ref{tab:evol}. For the sake of clarity, a selection of these properties is also presented in Fig.
 \ref{fig:panelevo}.

The general behavior we obtain for models leading to UMCO WD reproduces that described in Dom96. In
particular, the  lifting effect  of rotation  in the  core leads to a slower increase in the He-burning shell temperature, thus to a slower expansion and decrease in the H-burning shell temperature, and ultimately to a delaying in the second dredge-up (SDU), which allows a considerably higher CO core growth.    
%leads  to a delaying in the onset of the SDU, yielding a larger CO core.
%reduces the release of gravitational energy. As a consequence, the He-burning shell increases its temperature more slowly, the H-burning shell fades away later, and thus the second dredge-up (SDU) is delayed, which allows a considerably higher CO core growth. 
This is inferred  from Table \ref{tab:evol} and from  the top left panel of Fig. \ref{fig:panelevo}, which  shows the core 
mass at  the SDU in terms  of the initial mass  for different rotation
parameter values.  We note also the  resulting  larger core  masses with  increasing
$f$ (top right panel of the same figure). Because of  the reduction in maximum temperature  induced by core
rotation, carbon burning is prevented and the mass  of the resulting CO
core will  be larger than that
at which carbon  burning is expected in the absence  of rotation. As a
result, the mass  of the degenerate CO core will  be larger than 1.05  $M_{\odot}$ 
before the TP-AGB. By comparing the mass of the H-exhausted core at the end of core He-burning ($M_{\rm c,CHeB}$), and the core mass at the maximum advance of the SDU ($M_{\rm c,SDU}$) we can see that rotation even hampers the occurrence of the SDU in our 5  $M_{\odot}$  models. For a comparison,  the minimum initial mass for the occurrence of the SDU in nonrotating solar metallicity stars is $\approx$ 4  $M_{\odot}$ (see, e.g. \citealt{becker1979, boothroyd1999}).

The higher WD masses resulting from rotation are in line with recent observational constraints provided by the semi-empirical initial-final mass relation \citep{2019ApJ...871L..18C}. These authors 
%have shown that rotation leads to higher-mass WDs, 
attribute most of the observed scatter of the initial-final mass relation at high initial masses to the occurrence of rotation. Finally,  we  have neglected the effects of rotational mixing  during the evolution prior to the AGB phase, starting from the main sequence. This is not a minor simplification, which is expected to lead to higher core masses at the onset
of the AGB \citep{2019ApJ...871L..18C}.  As a result, the same  final core masses can be reached with a lower rotation parameter $f$ on the AGB.

%Since our interest in this work
%is not focused on the shape of  the initial-final mass relation, but %instead on the possibility of forming  UMCO WDs irrespective of the %initial mass of the progenitor star, this simplification does not bear %relevant consequences for the main conclusions of this work.

The occurrence of larger H-exhausted cores in rotating models leads to more luminous host stars (see also Dom96). Fig. \ref{fig:panelevo} shows how the peak luminosities during the TP-AGB increase with the rotation parameter $f$. This leads to more efficient stellar winds, to a shorter duration of the TP-AGB phase and to a decrease in the number of thermal pulses for faster rotating models. 
As in Dom96, we also obtain that the decrease in effective gravity of the He-exhausted cores of our models also affects the active burning shells and envelope structure, by making them more extended and cooler. Specifically, as $f$ increases, the maximum temperature at the base of the convective envelope ($T_{\rm BCE}$) decreases. Temperatures above  $T_{\rm BCE} \gtrsim 3\times 10^7$ K  leads to hot-bottom burning (HBB) \citep{boothroyd1991,ventura2005}. 
%which allows H-burning to actually occur at the base of the convective envelope. 
Its main nucleosynthetic effects reflect the occurrence of the CNO-cycle, that is, an increase in surface He and $^{14}$N, and a decrease in $^{12}$C and, to a minor extent, in $^{16}$O. The $T_{BCE}$ of our 5  $M_{\odot}$  models computed with rotation are below or just near the threshold for the occurrence of HBB. This is reflected in the high surface $C/O$ values at the end of their evolution, compared to those of more massive models (see Table \ref{tab:evol}). Indeed, in 5  $M_{\odot}$  rotating models, $^{12}$C is not destroyed by HBB but efficiently enhanced by the third dredge-up (TDU) episode described below. 

Strong HBB,  characteristic of intermediate-mass stars hosting the most massive cores, has been pointed to as responsible for the cessation of core growth \citep{poelarends2008}. However, this effect would be strongly dependant on the modeling of convection. Even though our models hosting the most massive cores do experience more modest core growth during their TP-(S)AGB than their less massive counterparts, we still identify core growth during interpulses. Therefore we consider that limited overall core growth should be ascribed to shorter TP-(S)AGB lifetimes, combined with efficient TDU.          

All our model stars experience the TDU, in which the base of the convective envelope advances inwards, and dredges-up matter previously synthesized in the convective regions associated to thermal pulses. The TDU efficiency is measured by the $\lambda$ parameter \footnote{$\lambda$ is the quotient between the core mass dredged-up after a thermal pulse and the core mass increase in the previous interpulse period (see, e.g. \citealt{mowlavi99}). High $\lambda$ values, that is, a very efficient TDU episode, may significantly hamper core growth during the TP-AGB phase.}. Its main nucleosynthetic effects are an increase in $^{4}$He and $^{12}$C and, to a minor extent, an increase in $^{16}$O. The maximum values of $\lambda$ tend to
decrease when $f$ increases, although differences become minor for our most massive models 
(M$_{ini}$ $\geq$ 7.5  $M_{\odot}$). The lower efficiency of the TDU with increasing $f$ is related to the relative weakness of the thermal pulses of models with rotation, which, themselves, is caused by their He-burning shell structures being more extended and cooler. 
For instance, the maximum thermal pulse luminosity, $L_{He}$, for our 7  $M_{\odot}$  models computed with $f$= 0, 0.1, 0.15 and 0.20 are, respectively, $1.59\times 10^8$, $1.15\times 10^8$, $8.33\times 10^7$, and $7.34\times 10^7$  $M_{\odot}$. 
In models without rotation, or in our massive rotating models (those whose initial mass is $\gtrsim$ 7.5  $M_{\odot}$), $\lambda$ tends to decrease with increasing initial mass. The reason is that more massive cores are more compact and hotter, $P_{rad}$ is more important in their He-burning shells, and thus degeneracy and thermal pulse strength consequently lower. Because the strength and duration of the instability decreases, the maximum depth of the subsequent TDU episode is also diminished  \citep[see][]{straniero2003}. 
%Note also that for our 7.5 and 8  $M_{\odot}$  models, in which peak $L_{He}$ is always greater than $1.59\times 10^8$  $L_{\odot}$ , a saturation effect is reached, and $\lambda$ dependency with rotation parameter becomes very weak.      

The interpulse period of our models (Table \ref{tab:evol}) does not always follow the trend mentioned by \citet{dominguez1996} for their 6.8  $M_{\odot}$  nonrotating model. These authors explain that, because of the lower H-burning shell temperatures of rotating models, the critical He mass for the thermal pulses to occur is reached later, and thus the interpulse period increases with $f$. In our case, models of initial mass $\leq$ 7  $M_{\odot}$  present less efficient TDU for higher $f$ cases, which would lead to somewhat more massive He buffers at the beginning of the interpulse. This would, at least partially,  balance the need for longer interpulse periods due to the less efficient H-burning shell. In addition, the similarly high $T_{\rm BCE}$ and $\lambda$ values of our most massive models lead to similar interpulse periods, regardless the considered $f$.

We have also explored the initial mass boundaries for the formation of white dwarfs of different types.
In Table \ref{tab:grid} we list the mass of the resulting
WD as  a function of  the initial stellar  mass $M_{\rm ini}$  and the
rotation parameter $f$.   Black, red, blue and  italic numbers denote,
respectively, the  evolutionary outcome of  normal CO, UMCO,  ONe, and
WDs  with  masses  exceeding  the Chandrasekhar  mass  value,  $M_{\rm
  Ch}$. We adopt  $M_{\rm Ch}= 1.37M_{\odot}$ that  corresponds to the
canonical Chandrasekhar mass expected for nonrotating stars.

%\citep{2004A&A...427..923C,2007A&A...465..249A}. 
%\begin{figure}
%    \centering
%    \includegraphics[width=1.00\linewidth]{m2du_fin.eps}
%    \caption{Core masses at the maximum inner advance of convection during the SDU episode (dashed line and squares) and expected WD masses (solid line and stars) versus initial mass for different values of the rotation parameter $f$. }
%    \label{fig:modmasses}
%\end{figure}

\begin{table}[t]
    \centering
    \begin{tabular}{cccccccc}%\toprule
    \hline
      $M_{\rm ini}/M_{\odot}$ & \phantom{abc} & \multicolumn{4}{c}{$f$} & \phantom{abc} & \\ \cmidrule{2-8}
    & 0 & 0.02 & 0.05 & 0.10 & 0.15 & 0.20 & 0.25 \\
    \hline
    \hline
    \noalign{\smallskip}
    4.0   & -  & - & - & - & 0.82 & 0.93 & 0.98 \\ 
    5.0   & -  & - & - & 1.02 & \umco{1.05} & \umco{1.11} & \umco{-} \\   
    5.5   &- & - & - & \umco{1.05} & \umco{1.14} & \umco{-} & \umco{-} \\ 
    6.0   & 0.90 & - & -    & \umco{1.08} & \umco{1.17} & \umco{1.29} & \umco{1.34} \\
    7.0   & 0.97 & 1.01 & \umco{1.05} & \umco{1.14} & \umco{1.24} & \umco{1.35} & {\it 1.48} \\
    7.5   & 1.02 & 1.03 & \umco{1.07} & \umco{1.15} & \umco{1.25} & \umco{-} & {\it -} \\
    8.0   & \one{1.08} & \umco{1.09} & \umco{1.12} & \umco{1.20} & \umco{1.29} & {\it 1.41} & {\it 1.56} \\
    8.5   & \one{1.16} & \one{1.17} & \one{1.18} & \one{1.25} & \one{1.35} & {\it -} & {\it -} \\
    9.0   & \one{1.23} & \one{-} & \one{1.36} & {\it 1.40} & {\it 1.48} & {\it -} & {\it -} \\
    9.5   & \one{1.35} & \one{-} & {\it -} & {\it 1.41} & {\it -} & {\it -} & {\it -} \\
      \noalign{\smallskip}
    \hline
      \noalign{\smallskip}
    \end{tabular}
    \caption{Resulting WD mass as a function of the initial mass $M_{\rm ini}$ and rotation parameter value $f$ for our computed models.   Black, red, blue and italic numbers denote, respectively, the evolutionary outcome of normal CO, UMCO, ONe , and WDs with masses exceeding the
    Chandrasekhar mass value for nonrotating stars. See also  Fig. \ref{fig:rotmodels}. %\san{he retocado el aspecto de la tabla para adaptarla al formato A&A. Por que hay casos que no estan calculados? Es por una razon fisica, lo de la w/wcritica?}.\PGP{Porque buscábamos los límites de outcome, y además había llenado varias veces mi disco duro :).}
    }
    \label{tab:grid}
\end{table}

\begin{figure}[t]
    \centering
    \includegraphics[width=1.0\linewidth]{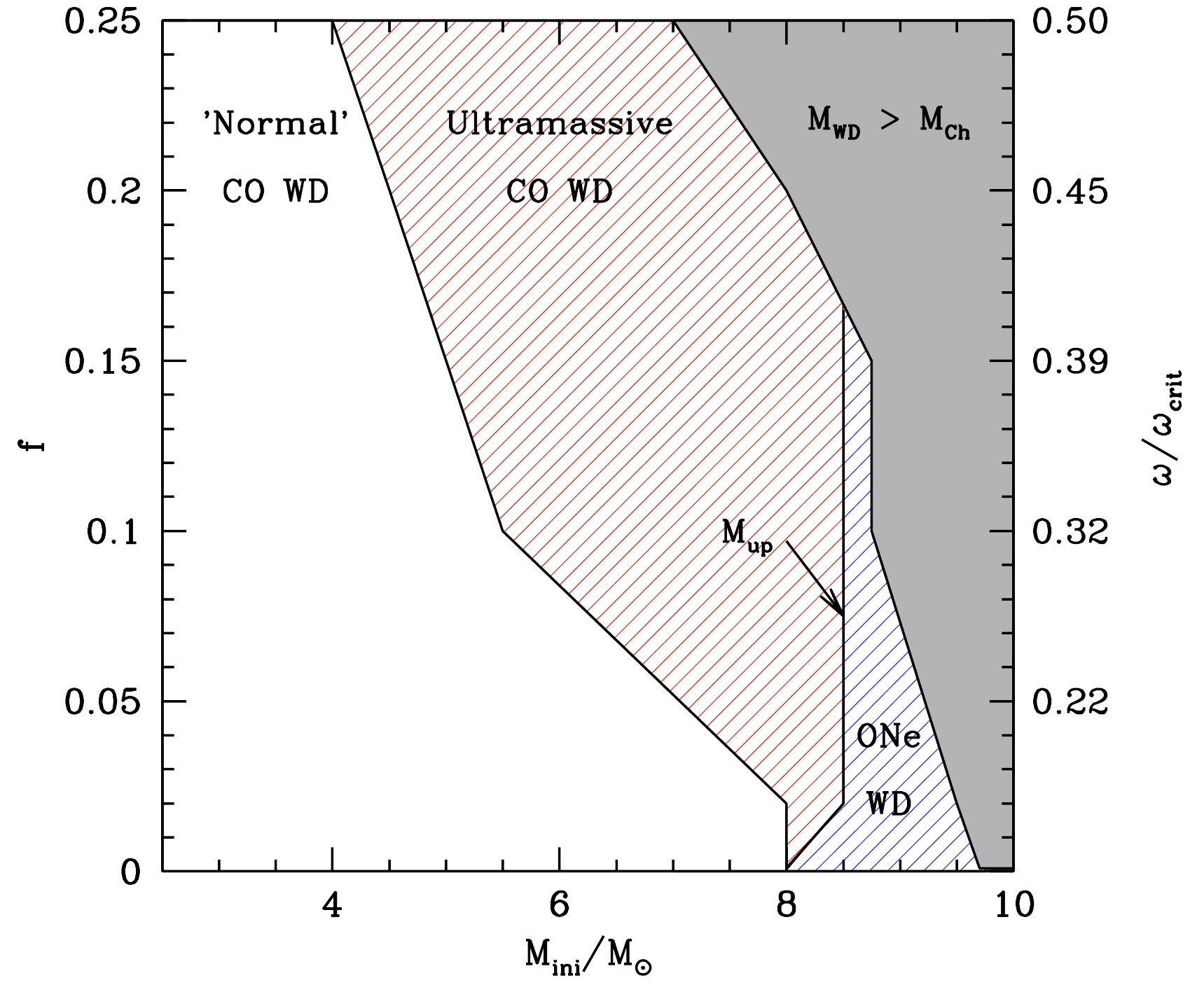}
    \caption{Evolutionary outcomes for the formation of normal CO, UMCO, and ONe WDs. Rotation parameter $f$ (and equivalent $\omega/\omega_{\rm crit}$) versus initial mass. The adopted value for the Chandrasekhar mass, $M_{\rm Ch}$,  corresponds to the canonical value expected for nonrotating stars.  $M_{\rm up}$ gives  the minimum initial mass threshold which leads to the onset of extended carbon burning.
   }
    \label{fig:rotmodels}
\end{figure}

 Fig.   \ref{fig:rotmodels}   illustrates   the  resulting   WD   core
 composition for the  set of $f$ values and initial  stellar masses we
 considered.  Even  for the lowest $f$  value of 0.02, we  obtain that
 the effects of rotation are noticeable and that the formation of UMCO
 WDs is possible.
%may be formed for an initial mass range between 8.5 and 9.25 \msun{}. 
The  range  of initial  masses  leading  to  UMCO  WDs widens  as  $f$
increases,  whereas  the  range  for the  formation  of  ONe-core  WDs
decreases significantly.
%\san{yo añadiria la siguiente frase porque es interesante tambien verlo desde el punto de vista de la $M_{ini}$: In other words, initial masses below $8\,M_{\odot}$ will lead to the formation of normal CO or UMCO WDs. Initial masses in the range $8-9\,M_{\odot}$ will form UMCO WDs unless the rotation parameter is lower than 0.02, where a ONe WDs is form. Initial masses greater that $9\,M_{\odot}$ will produce ONe WDs regardless of the rotation parameter.} \san{Punto y a parte}Note that our calculations lead to no ONe WD below the canonical Chandrasekhar mass for values of $f \geq 0.17$.
Even though the search for $M_{\rm  Ch}$ values for rotating models is
beyond the scope  of the present work,  we would like to  note at this
point  that  rotation is  expected  to  alter  the values  of  $M_{\rm
  Ch}$.  \citet{anand1965} used  Chandrasekhar's series  expansions to
show    that    rotation    should   increase    $M_{\rm    Ch}$    by
2$\%$. \cite{ostriker1968} built  axisymmetric differentially rotating
WD  models  which  could be  stable  up  to  masses  as high  as  $4.1
M_{\odot}$ More  recently, using 2D models,  \cite{yoon2005} confirmed
that  the critical  masses  for thermonuclear  or electron-capture  SN
explosions  was  expected  to  significantly exceed  $M_{\rm  Ch}$  in
rotating CO cores.

\begin{figure}
    \centering
    \includegraphics[width=1.0\linewidth]{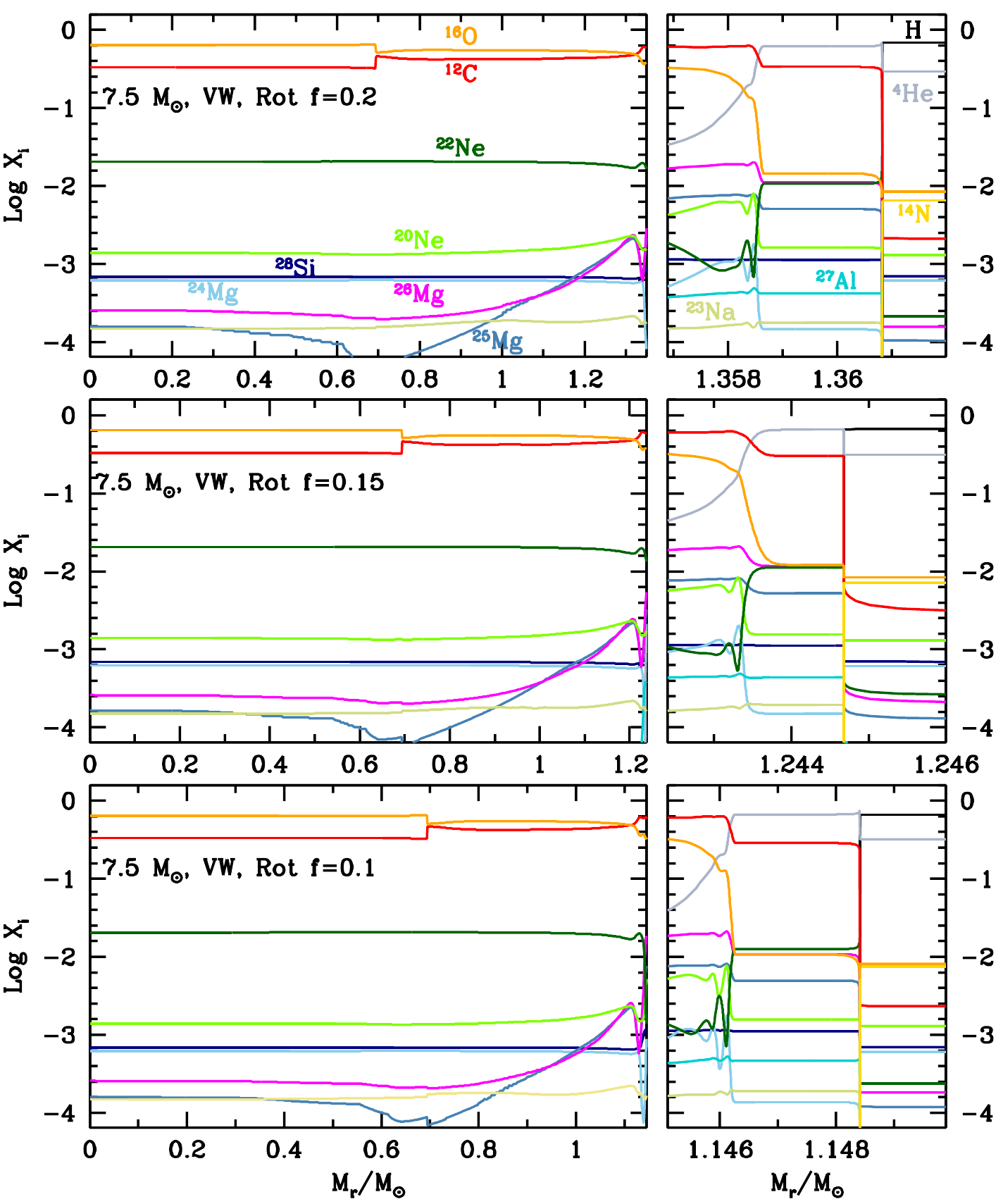}
    \caption{Inner chemical abundances at the TP-AGB for our $7.5 M_{\odot}$ initial model for different rotation
    parameters $f$  and VW mass-loss rates.}
    \label{fig:chem_m75}
\end{figure}

For  illustrative purposes,  we  show in  Fig. \ref{fig:chem_m75}  the
internal chemical  profiles of our  model star  of 7.5 \msun{}  at the
TP-AGB for different  rotation parameters $f$.
%The  occurrence of core
%rotation prevents  carbon ignition  from occurring, thus  allowing the
%formation  of  UMCO  WDs  of masses  well  above  $1.05M_{\odot}$.
%In general, the chemical  structure is not very different from  that of a
%typical  AGB star  of lower  mass.
The  three models  shown have  been
extracted  from the  last He  flash, so  the expected  pure He
buffer  below  the H-rich  envelope  has  been diluted  by  the
pulse-driven convection  zone, which  in turn  is responsible  for the
formation of the intershell rich in He and carbon.

\subsection{Mass-loss scenario}
\label{mass loss}

%, at which point the remnant was forced to abandom the TP-AGB.

According to existing literature,  intermediate-mass stars
develop off-centre carbon-burning when  their degenerate CO cores have
a mass $M_{\rm core}\gtrsim 1.05\,M_{\sun}$.  The minimum initial mass
for  this to occur ranges between  approximately
6 and  9\,$M_{\sun}$, see \cite{doherty2017}. The
degenerate CO  cores hosted  by stellar models  of initial  mass below
this threshold can still grow beyond the 1.05\,$M_{\sun}$ limit during
the TP-AGB;  however, this  growth is typically  limited to  values of
only $\sim 0.01\,M_{\sun}$  for models of solar  and Magellanic Clouds
metallicity, see  also Fig. \ref{fig:panelevo}. The  reasons for this
limited  core growth  during the  TP-AGB  are mainly  the strength  of
stellar winds and  the efficiency of the TDU. The initial
mass  of  the core  at  the  beginning  of  the TP-AGB  is  determined
essentially  by  the efficiency  of  the  SDU,  which depends  on  the
treatment  of convection  and  to  a minor  extent  on the  convective
boundary mixing, see \cite{2020MNRAS.493.4748W}.

Here, we  explore the possibility that  UMCO WDs could be  formed as a
result of a slow  growth of the CO core during  the TP-AGB. This could
be possible as long as the  minimum CO-core mass for the occurrence of
carbon  burning  is not  reached  before  the
TP-AGB.  Indeed,  during the TP-AGB,  core temperature, that  has been
increasing since the  end of core He burning,  starts to decrease,
and hence the onset of carbon burning that produces a ONe core is
avoided. We  carry out this experiment  by focusing on a  reduction in
the mass-loss rates.  Specifically, we  have followed the evolution of
an initially  7.8 $M_{\sun}$  model from the  ZAMS to  advanced stages
when  the CO  core mass  has grown  above 1.05\,$M_{\sun}$  during the
TP-AGB.

The efficiency of the TDU is critical for the core growth \citep[see][and references therein]{marigo2020} and, ultimately, for the eventual formation of an UMCO WD. Unfortunately, our knowledge of the TDU suffers from uncertainties derived from poorly known input physics, namely, the mixing and the determination of convective boundaries. 
In order to assess the latter effect, we calculated evolutionary sequences for a 7.8 $M_{\sun}$  initial model using two different prescriptions for the determination of convective boundaries. First we used the search for convective neutrality (SCN) approach \citep{castellani1971, frost1996}. The implementation of the SCN has been successfully used in the study of intermediate-mass stars in a wide range of masses and metallicities (see, e.g. \citealt{2014PASA...31...30K,doherty2017} and references therein).
The related algorithm aims to limit the effects of the sharp (and unphysical) discontinuity of the radiative gradient at the convective boundaries and, in practice, it works as an induced overshooting. As a consequence, it favors very efficient TDU and, thus, hampers rapid core growth in TP-AGB models.

We have also used the Schwarzschild criterion to determine convective boundaries. Naturally, TDU is less efficient and thus core growth and the formation of UMCO WD models are favored. Table \ref{tab:evol} summarizes our main results regarding the effects of stellar winds, which are further discussed below,) and those of the determination of convective boundaries. The TDU efficiency, represented by $\lambda$, is significantly higher when the SCN is used. When we compare the model computed with standard \citet{vassiliadis1993} wind rates and SCN, and the model computed with the same wind prescription and the Schwarzschild criterion (Schw) for convective limits, we see a decrease in $\lambda$ of $\approx$ 36$\%$ . It translates in a more massive final H-exhausted core, higher peak luminosities, more efficient stellar winds, and a shorter TP-AGB duration. As a consequence, an UMCO WD can be obtained with a factor 2 decrease in VW mass-loss when Schw is used, whereas a decrease in wind rates of a factor slightly lower than 10 is required when SCN is implemented. From the nucleosynthetic point of view, the less efficient TDU obtained with Schw, together with highly efficient HBB, leads to a lower surface abundance $C/O$ ratio.

%\begin{figure}
%        \centering
%        \includegraphics[width=1.0\columnwidth]{quimica-iniciales.eps}
%        \caption{Internal distribution of mass fraction of selected chemical %elements in terms of the outer mass coordinate  corresponding to $1.156 %M_{\odot}$ultra-massive WD models resulting from the various evolutionary 
%        scenarios considered in this paper. From top to bottom it is displayed: 1) %ONe WD model from \cite{2019A&A...625A..87C}, 2) CO WD model formed as a result of %reduced mass loss, 3) CO WD model implied by rotation, and 4) CO model resulting %from the merger of two equal-mass WDs. All of the chemical profiles are shown at %the beginning of the cooling track before the occurrence of the mixing process in %the core as a result of the inversion of the mean molecular weight {\bf ojo el %piojo, hay que agregar el merger en el ultimo panel}.} 
%        \label{perfil-pilar.eps}
%\end{figure}

AGB winds are probably caused by pulsation-enhanced radiation pressure
on  carbonaceous   or  silicate   grains  present  in   extended  cool
envelopes. Collisional coupling between grains and gas allows this gas
to be ejected together  with grains. Recently, \cite{2019A&A...626A.100B} have  studied wind formation and
 the  properties of  stellar winds  for different  stellar masses  and
 luminosities appropriate for the AGB phase by modeling the mass-loss
 process  from  first  principles,  in  particular  the  inclusion  of
 frequency-dependent  radiation-hydrodynamics   and  a  time-dependent
 description of dust condensation and evaporation.
%These detailed simulations
%show that mass-loss rates strongly correlates with $L$ and with the ratio $L/M$, predicting an increase in the mass-loss rates with
%$L/M$ substantially steeper than given by the \citet{reimers1975} mass-loss prescription.
Their  derived mass-loss  rates  show a  large  dispersion with  input
parameters such as stellar luminosity,  stellar mass, the abundance of
seed particles, grain size and gas-to-dust mass ratio.

%\citet{reimers1975} mass-loss prescription was derived for supergiants, by matching spectra of binary %systems consisting of a red supergiant and an unevolved companion.
%Both Blo95 and Sch05 are modifications of Rei75 formula. Blo95 introduced the effects of pulsational %models of Mira variables, whereas Sch05 derived their expression by assuming that AGB winds are result %from the spillover of the extended chromosphere due to the action of waves. VW93 aimed to match the %pulsation period of observed red giant and AGB stars. 

%While the use of the different mass-loss prescription implemented in our code yields different TP-AGB durations and thus different final degenerate core masses, we find that the formation of UMCO WDsrequire reduction in the mass-loss rate prescription by factor of 10, when using the mass-loss prescription by \citet{vassiliadis1993}, and $\eta \sim 10^{-4}$ when using the prescription by \citet{bloecker1995}. Note that a frequently used $\eta$ value in this case is 0.02, which was derived from comparisons to observations of Li-rich giants in LMC \citep{ventura2000}. 
%Our mass-loss rates for the computed evolutionary sequence is given in Fig., which can be compared with the predictions by \cite{2019A&A...626A.100B} ............De que orden es el maximo Mdot para obtener una UMCO ? En funcion de lo que de esta comparacion quiza podriamos concluir cuan lejos de la realidad estaria este escenario. Es que la disperson del Mdot que muestra Bladh es muy grande}.

\begin{figure}
  \centering
    \includegraphics[angle=-90,width=1.0\linewidth]{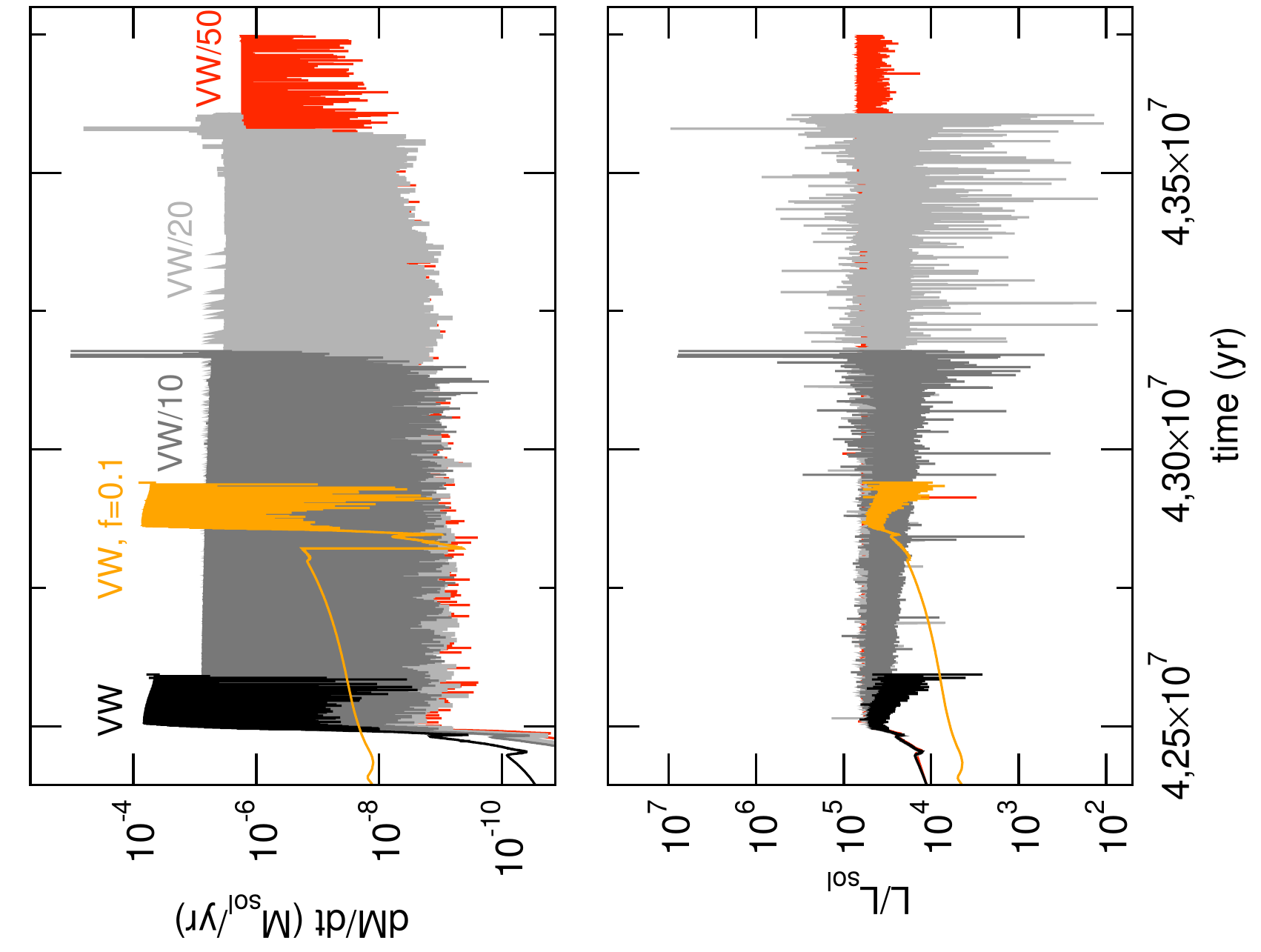} %el eps esta rotado
    \caption{Mass-loss rates and surface luminosities versus time for models computed with standard VW mass-loss rates (black), standard VW over 10 (dark grey), standard VW over 20 (light grey), standard VW over 50 (red), and standard VW and rotation with $f= 0.1$ (yellow).}.
    \label{fig:mdots}
\end{figure}

Usually, wind prescriptions in stellar  evolution codes are the result
of  semiempirical approaches.   The mass-loss  rates for  some of  our
computed evolutionary sequence are given in Fig. \ref{fig:mdots}.  All
the sequences\footnote{Some of our sequences  reach, toward the end of
  the TP-AGB phase,  unrealistic high value in  the surface luminosity
  caused  by numerical  artifacts due  to the  dominance of  radiation
  pressure at  the base  of the  star envelope.},  except for  the one
calculated with standard VW rates, lead to UMCO WDs. Specifically, the
cases in which standard  rates are reduced by a factor  10, 20, and 50
yield, respectively, degenerate CO-core  masses of $1.066, 1.113$, and
$1.165\,M_{\sun}$. The case in which  mass-loss rates are reduced by a
factor 5 leads  to a degenerate core of  $1.044\,M_{\sun}$, very close
to the  theoretical mass limit  of UMCO  WD. 
The summary of the main characteristics of our sequences computed with different wind rates is shown in Table \ref{tab:evol}.  We mention that 
our reduced mass-loss rates lead to substantially longer AGB lifetimes (except
when  Schw is used, in which case AGB lifetimes are substantially lower)
than those provided by observational 
constraints that support short TP-AGB lifetimes for stars as massive as 
$4-5\,M_{\sun}$, (\citealt{2007A&A...462..237G}, \citealt{2019MNRAS.485.5666P}), thus casting doubt on the use of low
mass-loss rates prescriptions with efficient TDU. However, it should be mentioned that these authors extend their analysis up to initial masses of about $5\,M_{\sun}$,
and discard more massive progenitors, which are the ones in which we are interested in our simulations, i.e, progenitors with initial masses larger than $7.5\,M_{\sun}$.

The comparison
with  \cite{2019A&A...626A.100B}  is  not  straightforward,  as  these
authors focused  on stars of  initial masses up to  $3\,M_{\sun}$, and
used a  method to derive  mass-loss rates  different from that  of VW.
The  latter,  although relatively  simple,  has  proved successful  in
reproducing  luminosity tips  of  Magellanic Cloud  stars, or  surface
abundances of AGB stars up  to $\sim$ 5 \msun\ \citep{vassiliadis1993,
  lattanzio2016}. However,  as noted by  \citet{hofner2018}, empirical
mass-loss rates expressed  as functions of specific  variables must be
applied with caution, as those variables are affected by other stellar
properties   and    observational   biases.    \citet{groenewegen2009}
investigated the relation between mass-loss rates, luminosity and dust
temperature for  an extended sample  of LMC  and SMC stars.  They also
adapted their synthetic evolution  code to compare theoretical results
to  observations,  and  to  extend  the  applicability  of  VW  up  to
$7.9\,M_{\sun}$ models. They concluded that, in general, the mass-loss
prescription by  VW worked  well, but pointed  to deficiencies  in the
high  $\dot  M$  range.  A  recent revision  of  the  latter  work  by
\citet{groenewegen2018} yielded similar results.

Mass-loss  determination methods  are
derived and/or tested through observations. Given the standard initial
mass function  and the shortness of  the duration of the  most massive
TP-AGB stars, related observations are likely  to be far from the mass
range in which  we are interested in the present  work. Mass-loss rate
uncertainties may thus be higher for the high mass range of AGB stars.
%The trend reported by \citet{2019A&A...626A.100B}, that mass-loss rates increase with $L/M$, probably holds beyond the 3 \msun mass threshold of their work, as it is a reflect of the importance of the gravitational potential well. This might support the case that mass-loss rates might be overestimated for the most massive AGB and Super-AGB stars. NO ESTOY MUY SEGURA DE ESTO ÚLTIMO, LAS RECETAS DE PÉRDIDA DE MASA YA TIENEN EN CUENTA EL EFECTO DE LA MASA. LA CUESTIÓN ES MÁS BIEN SI ES SUFICIENTE. SERIA ESTUPENDO UN TRABAJO COMO EL DE BLADH, PARA MASAS MAYORES, Y COMPARANDO CON FÓRMULAS SEMIEMPÍRICAS BASADAS EN RELACIÓN PERIODO-LUM... SIGH...
A  recent study  by \citet{decin2019}  points  to the  fact that  mass
loss-rates  in OH/IR  stars  might  be overestimated  up  to a  factor
100. According to these authors, the presence of undetected companions
in extreme OH/IR stars, for  which binary frequency is 60-100$\%$, may
cause episodes  of equatorial  density enhancements which  might mimic
the effects  of extreme  superwinds. In  addition, these  authors show
that  the analysis  of low-excitation  CO lines  is the  most reliable
method to estimate mass-loss  rates in (extreme) OH/IR-stars, yielding
typical       values      in       the      range       (0.1-3)$\times
10^{-5}\,M_{\odot}$/yr.  Our model stars
computed with the standard VW experience on average winds between 4 and 7 $\times
10^{-5}$\msun{}/yr   during    most   of   their   AGB    phase. Thus,  they are good candidates to  be cases of overestimated mass-loss rates.

%Thus, the  authors fix  the single-scattering
%radiation pressure limit (and even a lower limit by a factor of a few)
%as an upper threshold for mass  loss rates.  
%Note that our model stars
%computed with the standard VW experience winds between 4 and 7 $\times
%10^{-5}$\msun{}/yr   during    most   of   their   AGB    phase   (see
%Fig. \ref{fig:mdots}). Thus,  they are good candidates to  be cases of
%overestimated mass-loss rates.

Mass-loss rates  lower than expected  according to current  models may
not  only  help to  solve  the  discrepancy  between the  duration  of
superwinds (from observations) and the  time required for the stars to
become WDs  \citep{decin2019} but  also the unexpectedly  high masses
($\gtrsim 0.1\,M_\odot$) accreted by  C-enhanced metal-poor stars from
their companions  reported by  \citet{abate2015}, the survival  of gas
planets  close  to  their  host  stars  \citep{villaver2007},  or  the
posibility that planets might shape the planetary nebula of their host
star \citep{hegazi2020}. Moreover, indication of limitations in 
how standard models address mass loss and TDU efficiency at higher masses 
is provided by the semi-empirical initial-final  mass relation that predicts
a systematic offset of $\sim 0.1\,M_\odot$ more massive WDs than theoretical models, see \cite{2018ApJ...866...21C}.

%{\bf We are aware that our reduced-mass loss scenario  
%is the weakest one  we propose to create UMCO WDs. Admittedly, the reduction
%in the mass-loss rates we need to form UMCO WDs appears quite drastic, albeit
%not entirely discarded by different pieces of observational evidence, as we mentioned. Be that %as it may, we insist on that it constitutes an alternative
%scenario to identify possible effects of single progenitor
%evolution on the resulting WD chemical profiles, and the impact on its further
%evolution and observable consequences.}

 We are aware that our reduced-mass loss scenario  
is the weakest one to create UMCO WDs. However, this scenario
cannot be entirely discarded by different pieces of observational evidence, as we mentioned, including the persisting uncertainties in the efficiency
of TDU. Be that as it may, we insist on that it constitutes a plausible
scenario to identify possible effects of single progenitor
evolution on the resulting WD chemical profiles, and the impact on its further
evolution and observable consequences.

%\PGP{Important \citet{decin2019}. Occurrence of CEDE mimic short episodes of very high MLR. Reported MLR have been overestimated by a factor 100. VW93 should be safe because by construction always below single scattering limit. Tested by Groenewegen and Sloan 2009, but if MLR overestimated in observations, prescriptions may end up overestimating too. Check in code how vinf for single scattering limit is obtained.}. 

%\PGP{Neither binarity in non-degenerate stages can be claimed as an option  for the formation of UMCO. Indeed, even if a primary %(mass-losing) star ejects its H-rich envelope during a mass transfer episode, either during its main %sequence or during its RGB, a subsequent SDU episode may still ensue, powered by a He-burning shell %\citep{gilpons2002}. It should be noted, though, that binarity typically shifts $M_{up}$ up about %1.5\msun{}. Considering reasonable convective boundary mixing variations may alter also shift $M_{up}$, %in this case downwards, but it  does not major consequences (after considering this shift), neither on %the efficiency of the SDU, or on the core mass thresholds which lead to C-burning. }

\subsection{Double-WD merger scenario}
\label{binary}

 Ultra-massive WDs can also be the  result of binary evolution channels, mostly
 the merger of two intermediate-mass CO-core WDs (WD+WD merger) with a
 combined     mass    below     the    Chandrasekhar     limit,    see
 \cite{2020A&A...636A..31T}.  By using  a  detailed smoothed  particle
 hydrodynamics  code,  \cite{2014MNRAS.438...14D} have  presented  the
 merger  properties   of  WD+WD  systems  and   investigated  how  the
 components mix chemically.  They find that systems with  a mass ratio
 close to unity yield an almost  spherical remnant and that the mixing
 between the components is maximum.

In this study we have considered  the formation of an ultra-massive WD resulting from  the  merger  of  two  equal-mass WDs, by  assuming  the  extreme situation of  complete mixing  between the  two WDs and a CO core
for the merged remnant.   To this  end we
adopt the internal  distribution of carbon, oxygen, and  He of the
0.58\,$M_{\sun}$ WD model  computed in \cite{2016A&A...588A..25M} from
the  evolutionary history  of its  progenitor star.  Assuming complete
mixing  between both  WDs, the  mass  fraction of  He, carbon  and
oxygen throughout  the interior of  our UMCO WD model  results $X_{\rm
  He}=0.017$, $X_{\rm C}=0.363$ and $X_{\rm O}=0.62$.  We mention that
our interest in this study are the advanced evolutionary stages of the
WD and  not the  evolutionary stages shortly  after the  merger event,
which cannot be adequately followed by our stellar evolution code.
Finally, we have not considered in this work the possibility that WD mergers lead to ultra-massive ONe-core WDs, as predicted in \cite{2020arXiv201103546S}. The reason is that we do not expect such ONe WDs to present marked differences in their pulsational properties as compared
with the  ONe-core models resulting from single evolution that is explored in the present paper.

%leads our model to lose most of its envelope relatively quickly, so that the core growth during the TP-AGB varies from 1.015 \msun{} (at the early-AGB) to 1.024 \msun{} (at the end of our calculations). 
%We have performed a series of different tests by varying mass-loss rates in order to obtain ultra-massive white dwarfs. For the first test, we have considered 

%\begin{figure*}
%        \sidecaption
%\includegraphics[width=12cm]{lmdot.eps}
%        \caption{mdots} 
%        \label{perfil-merger.eps}
%\end{figure*}

 \section{Evolutionary properties}
\label{results}

\begin{figure}
        \centering
        \includegraphics[width=1.\columnwidth]{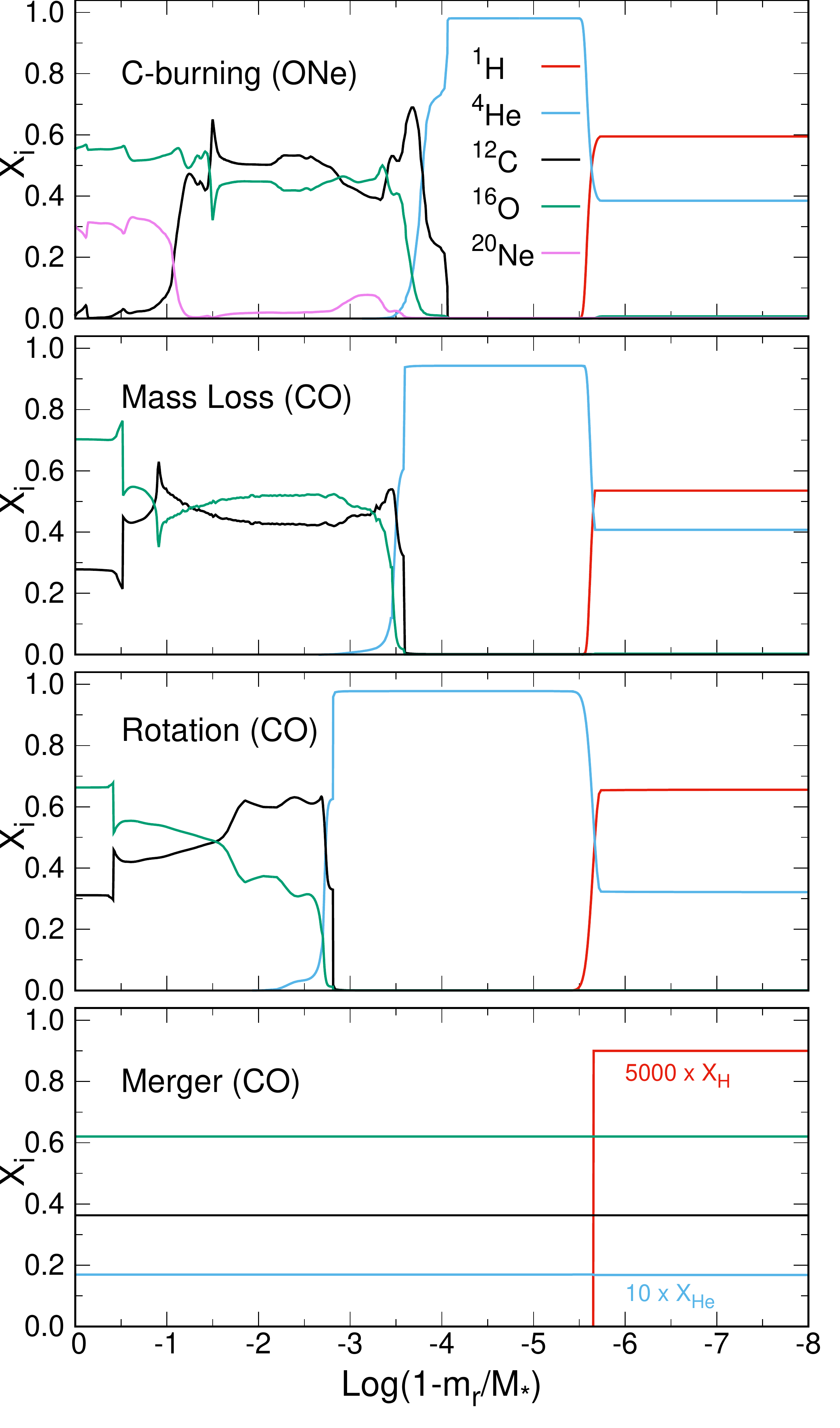}
        \caption{Internal abundance by mass of selected chemical elements versus the outer mass coordinate  for the  $1.156 M_{\odot}$ WD models resulting from the evolutionary 
        scenarios we studied. From top to bottom we show the ONe WD model from \cite{2019A&A...625A..87C},  the CO WD model resulting from reduced mass loss during progenitor evolution,  the CO WD model implied by rotation, and the merged CO model. Models are shown at the beginning of the cooling track before the core mixing induced  by  the mean molecular weight inversion.} 
        \label{perfiles-inicial.eps}
\end{figure}

In this section, we compare the evolutionary properties of our UMCO WD
models resulting from the two  single-evolution scenarios and from the
binary merger  we described  previously, with  those predicted  by the
ONe-core WD models resulting from off-center carbon-burning during the
single evolution  of progenitor star  \citep{2010A&A...512A..10S}.  To
isolate  the  impact of  the  internal  composition predicted  by  the
possible  scenarios,  we  have  adopted   the  same  stellar  mass  of
1.159\,$M_{\sun}$  for  all of  our  WD  sequences\footnote{We do  not
  expect  a qualitative  change  of  the conclusions  of  our work  by
  considering  other  mass  values  for the  ultra-massive  WDs.}.  In
addition we  have assumed  the same  H content  for all  of our
sequences, except for the merger case.  In particular, it was taken as
nearly  the maximum  allowed by  evolutionary considerations  for such
stellar  mass, $\log(M_{\rm  H}/M_{\star})\approx -6$.  This limit  is
imposed by the  occurrence of unstable nuclear burning  on the cooling
track  for   the  WD  mass   considered.  Finally,  the   ONe-core  WD
evolutionary    and    pulsational     models    were    taken    from
\cite{2019A&A...625A..87C} and  \cite{2019A&A...621A.100D}. Such models were
computed  with the  same codes  we use  to compute  the evolution  and
pulsations   of  our   UMCO  WD   models  (see   Sect.\ref{codes}  for
details). The evolution  of all of our WD sequences  has been followed
from very  high luminosities, through  the domain of the  pulsating ZZ
Ceti stars, down to very low surface luminosities.

In  Fig.  \ref{perfiles-inicial.eps}  we  display  the  internal  chemical
profile of  our UMCO WD  models  (see   Sect.  \ref{scenarios}  for  details),
together with the chemical profile of the ONe-core WD model studied in
\cite{2019A&A...625A..87C} (top panel).   Specifically, the second and
third panels illustrate, respectively, the chemical profiles resulting
from  reducing the  mass-loss rates  of an  initially $7.8  M_{\odot}$
progenitor star  and from considering  core rotation ($f=0.1$)  in the
AGB phase of an initially  $7.6 M_{\odot}$ progenitor star. The bottom
panel shows the  chemical profile we adopt for our merged WD.   The  chemical profiles  correspond  to
ultra-massive WD models  at the onset of their cooling  phase prior to
the onset of  element diffusion and before the core  mixing implied by
the  inversion  of the  mean  molecular  weight  is performed  in  our
simulations.

\begin{figure}
        \centering
        \includegraphics[width=1.0\columnwidth]{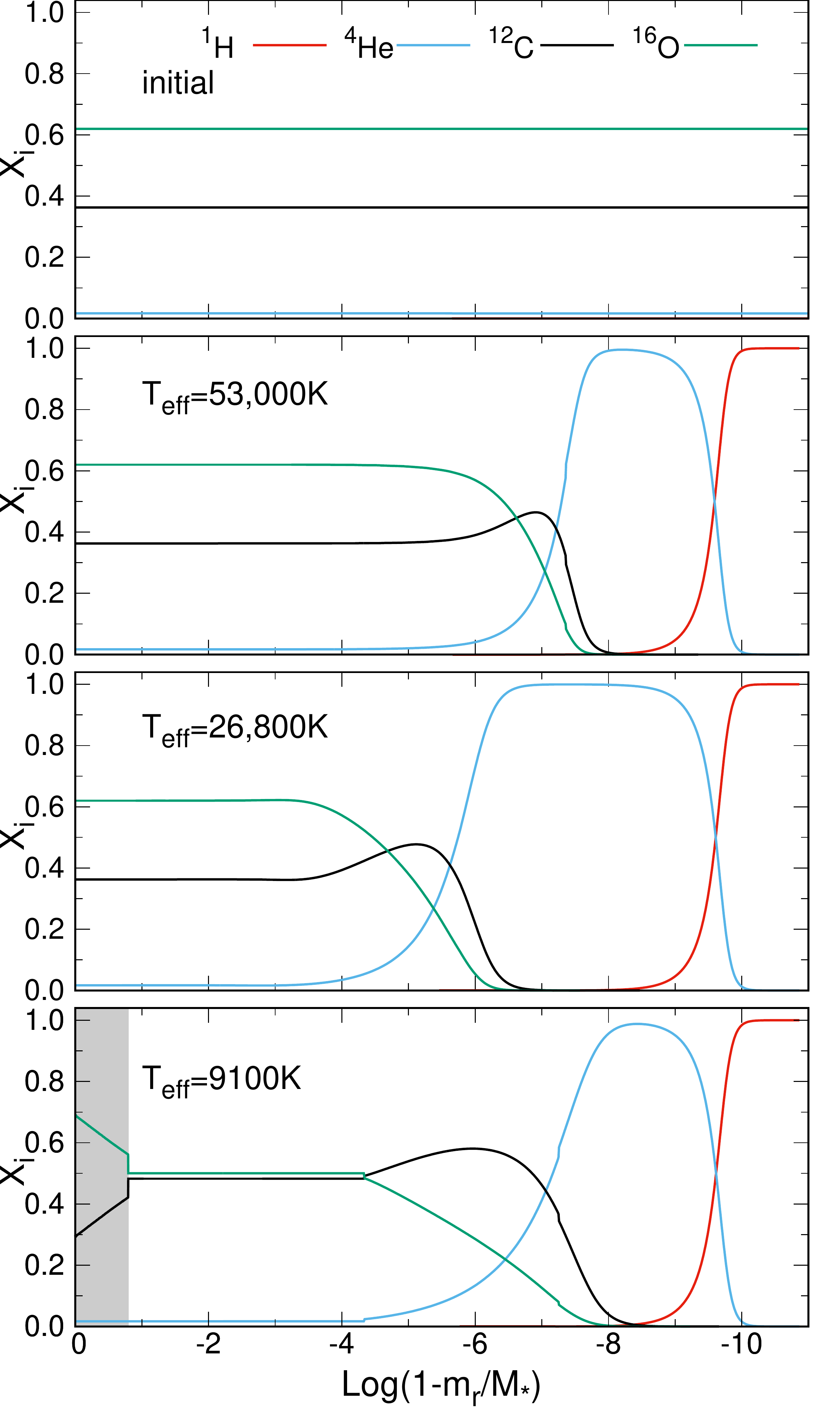}
        \caption{Internal abundance by mass  of H, He, carbon and oxygen  in terms of the outer mass coordinate  at various cooling stages of our $1.156 M_{\odot}$ UMCO WD sequence resulting from the merger scenario. Each stage is labeled by the effective temperature of the model. The gray area in the bottom panel
        indicates the domain of core crystallization. } 
        \label{perfil-merger.eps}
\end{figure}

The  chemical structure  expected in  ultra-massive WDs  is
markedly different  according to the evolutionary  scenario that leads
to their formation. This is true for both the core and the envelope.
%Not only WD+WD merger but also single stellar evolution can yield the formation of UMCO WDs,
%with chemical structures different from that predicted by progenitors that experienced carbon burning (top panel). 
The chemical profile that result in  the outer part of
the CO core is different in the mass-loss and rotation scenarios.
%The  reduction of internal temperatures and densities caused by the lifting effect of core rotation yields a quite different CO stratification of the core (see also Dom96). 
%Rotation  causes the AGB lifetime of progenitor to be larger and hence a consequent
%increase in the CO-core mass during the early AGB, thus affecting the CO stratification, as compared with the case of no rotation.  
The  He content of the UMCO WD  formed as a result
of  rotation  is larger  than  of  the  UMCO  WD formed  via  reduced
mass-loss rates  and of the  ONe-core WD.  This is because  the lower
temperature  and  pressure  prompted   by  core  rotation  favor  the
formation  of  a  less  degenerate  core  and  thus  a  larger  He
content. Since  we expect  that the internal  angular momentum  of the
core is conserved during the transition  from AGB to the hot WD stage,
we  assume  that no  He  is  burnt during  the  hot  stages of  WD
evolution.  We find  that allowing  He burning  to operate  in our
nonrotating WD models, the final He content of the WD is 
reduced. However,  if this were the  case, we find that  this reduction
of the He content does not change the conclusions of our
work.

%This is due to the larger degree of core degeneracy in the first two scenarios, resulting from the large number of thermal pulses  experienced by progenitor star, whilst in the case
%of rotation, the lower temperature and pressure prompted by core rotation favour the
%formation of a less degenerate core and thus a larger He content.  

The H content we adopt for our merged WD deserves some
comment. Because H is usually not included in smoothed particle
hydrodynamics simulations of double WD mergers, the amount of H
that    survives    the    merger   event    is    roughly    inferred
\citep[][]{2018MNRAS.476.5303S}.   It is  expected  that  most of  the
total  initial H  will be  destroyed by  the high  temperatures
during and  after the  merger \citep{2014MNRAS.438...14D}. In  view of
the lack  of consistent  estimations, we have  inferred the  amount of
H remaining  in our merged  WD in  a very simple  way: assuming
complete mixing of the two WDs (see Sect. \ref{binary}) and a H
content  of  $10^{-4}M_{\sun}$  for  each WD,  the  mass  fraction  of
H throughout  the 1.156  $M_{\sun}$ WD results  $X_{\rm H}  = 2
\times 10^{-4}\,M_{\sun} / 1.156\,  M_{\sun}$ = $1.73 \times 10^{-4}$.
At  the beginning  of the  cooling track,  H burning  occurs in
regions below log(1-$m_r/M_{\rm WD}$)=-5.6 in our 1.156\,$M_{\sun}$ WD
model. Assuming that all of the H below this layer is burnt, we
get   the   H   profile   shown  in   the   bottom   panel   of
Fig.\ref{perfiles-inicial.eps}.  The  remaining   total  H  content
amounts   to  $   5   \times  10^{-10}\,M_{\sun}$.   For  the   He
distribution, we have assumed no  He burning, despite some burning
is      expected       according      to       merger      simulations
\citep{2014MNRAS.438...14D,2016ApJ...819...94K}. We  stress again that
the  chemical  stratification  we   assumed  for  our  merged  remnant
corresponds to  the extreme situation  of complete mixing  between the
two WD  components. It is  beyond the scope  of this paper  to analyze
other  possible situations  that  would result  particularly from  the
merger      two     WD      remnants      of     different      masses
\citep{2014MNRAS.438...14D}.

Three main processes change the  internal chemical distribution of our
models: the mixing  of all core chemical components  during the pre-WD
and  onset  of cooling  as  a  result of  the  inversion  of the  mean
molecular weight, element  diffusion during most of  WD evolution, and
phase    separation    of     core    chemical    constituents    upon
crystallization. Fig.  \ref{perfil-merger.eps} illustrates  the impact
of diffusion and  phase separation for the UMCO WD  resulting from the
merger   scenario.   Diffusion   profoundly  changes   the   abundance
distribution  of the  initial  WD model  (top  panel). In  particular,
vestiges of inner  H surviving the merger episode  float to the
surface as  a result of  gravitational settling, thus forming  a thin,
pure  H envelope  at  rather high  effective temperatures.  The
formation of a pure He buffer  below the H envelope is also
predicted. At early stages, the  mass of this He buffer increases,
but as evolution proceeds chemical diffusion carries He to deeper
layers,  and   the  He   buffer  becomes  thinner   again  (bottom
panel). This  behavior can  be understood  as follows.  Generally, the
chemical  profile  in the  envelope  of  a  WD  is determined  by  the
competition between basically partial  pressure gradients, gravity and
induced  electric  field.   Diffusion  velocities are  given  by  (for
simplicity we neglect thermal diffusion)

\begin{equation}
\frac{dp_i}{dr}+\varrho_i g-n_iZ_ieE=
\sum_{j\ne i}^NK_{ij}\left(w_j-w_i\right),
\label{diff1}
\end{equation}

\noindent where, $p_i$, $\varrho_i$, $n_i$, $Z_i$, and $w_i$ are,
respectively, the partial pressure,  mass density, number density,
mean charge, and diffusion velocity for  chemical species  $i$, and $g$ the
gravitational acceleration \citep{1969fecg.book.....B}. $N$ is  the number of  ionic species
plus  electrons and $K_{ij}$ the resistance  coefficients. This set of equations is solved together with the equations
for  no  net mass  flow $\sum_iA_in_iw_i=0$ and no electrical current $\sum_iZ_in_iw_i=0$. Assuming 
only two ionic species  with mean charge 
$Z_{1}$ and $Z_{2}$ and atomic mass number  $A_{1}$ and $A_{2}$,
electron to have zero mass, and an ideal gas for the ions, then Eq. \ref{diff1} leads for the diffusion velocity
(positive velocity means that element diffuses upward)

\begin{eqnarray}
\lefteqn{{K_{12} f_n w_1}
=\left({-\frac{A_1}{Z_1}} + {\frac{A_2}{Z_2}} \right) 
m_{\rm H} g \ + \left({- \frac{1}{Z_1}} + {\frac{1}{Z_2}} \right) k_BT \ {{{\rm d}\ {\rm ln}\ T} \over {{\rm d}r}}}\nonumber
\\ & &
+\ {{\frac{k_BT}{Z_2}} {{{\rm d\, ln}\, n_2} \over {{\rm d}r}}\ -\ {{k_BT}
  \over{Z_1 }} {{{\rm d\, ln}\, n_1} \over {{\rm d}r}}},
\label{velohe}
\end{eqnarray}

\noindent  where $f_{n}$  is a  positive  factor that  depends on  the
number density and atomic charge of species and
%$f=  (1+A_1 n_1/A_2
%n_2) (n_1 Z_1 + n_2 Z_2) / n_1  Z_1 n_2 Z_2$. 
$m_{\rm H}$ the  H-atom mass. The first term  on the right-hand
side  of  Eq. \ref{velohe}  takes  into  account the  contribution  of
gravity  (and the  influence of  the  induced electric  field) to  the
diffusion velocity and  the second term gives the  contribution of the
temperature  gradient. These  two terms,  that are  usually
referred  to  as gravitational  settling,  cause  lighter elements  to
diffuse  upward. The  third and  fourth  terms refer  to the  chemical
diffusion contribution  resulting from gradients in  number densities.
In the  case of a  mixture of $^{4}$He and  $^{12}$C, the term  due to
gravity vanishes, and gravitational settling is thus strongly reduced,
resulting  solely  from  the   contribution  due  to  the  temperature
gradient, which is responsible for the formation (and thickening) of a
pure        He        buffer,        as        displayed        by
Fig.   \ref{perfil-merger.eps}.   However,    as   cooling   proceeds,
temperature  gradients in  the  He/carbon  transition zone  become
smaller and chemical  diffusion takes over, causing  He to diffuse
downward. This results in a reduction in the He buffer mass at low
effective temperatures.  In the  case of a  mixture of
H and He, the gravity term (first term) does not vanish; it
becomes  dominant  and  causes  H   rapidly  to  float  to  the
surface.  Finally,  the  bottom  panel shows  the  imprints  of  phase
separation on  the core  chemical composition  during crystallization,
with the  consequent increase  of oxygen  in the  solid core.  At this
stage, the mass of the crystallized core amounts to about 85$\%$.

\begin{figure}
        \centering
        \includegraphics[width=1.0\columnwidth]{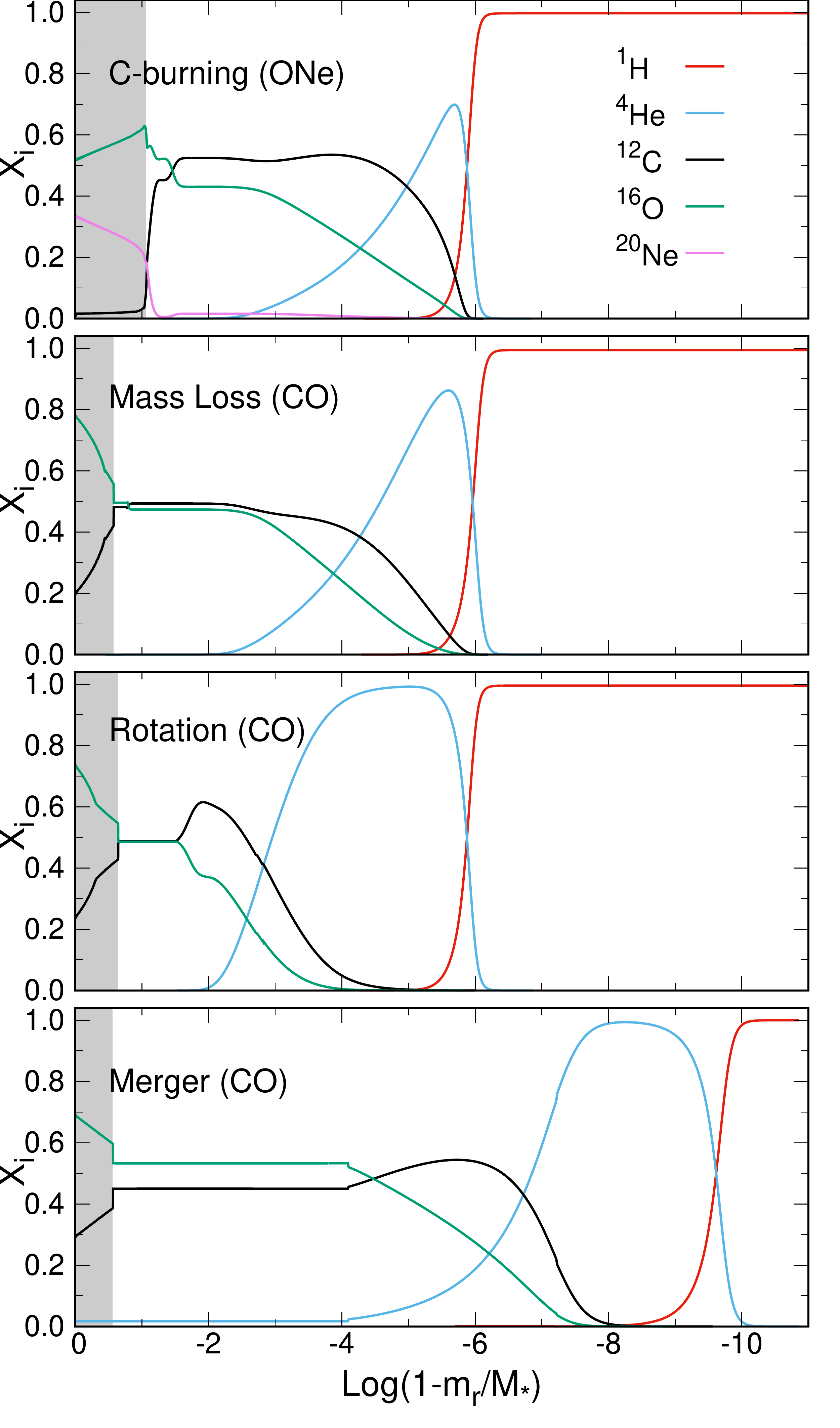}
        \caption{Same as Fig. \ref{perfiles-inicial.eps} but at an effective temperature
        of $T_{\rm eff} = 11,100$ K, representative of the domain of pulsating ZZ Ceti stars.  The gray area in each panel
        indicates the domain of core crystallization.} 
        \label{perfiles_zz.eps}
\end{figure}

The  initial chemical  profiles  of our  WD  models are  substantially
altered by  the time evolution has  proceeded to the domain  of the ZZ
Ceti instability  strip. This is shown  in Fig. \ref{perfiles_zz.eps},
where  it   is  displayed  the   internal  chemical  profile   of  our
ultra-massive  $1.156\,M_{\odot}$ WD  models resulting  from the  four
evolutionary  scenarios at  $T_{\rm  eff} =  11\,100\,$K.  In all  the
cases,  a  pure H  envelope  is  formed.  Also at  this  stage,
diffusion has  strongly smoothed out  the initial chemical  profile of
the  WD models,  a  relevant aspect  for  pulsational properties,  see
Sect.  \ref{pulsation_results}.  For  the  WD  models
resulting from the single evolution scenarios, the initial pure He
buffer  has  been strongly  eroded  by  chemical diffusion.   This  is
because  for such  scenarios the  He buffer  was formed  in deeper
regions (because  of the  larger H  envelopes expected  in such
cases), where the settling contribution resulting from the temperature
gradient becomes  smaller. Hence,  chemical diffusion  dominates, thus
causing  He to  diffuse downward,  and the  pure He  buffer to
disappear.  However,  we mention  that  in  our treatment  of  element
diffusion, we  have not included  the effect of Coulomb  separation of
ions \citep{2010ApJ...723..719C,2013PhRvL.111p1101B},  as a  result of
which  ions with  larger $Z$  should  move to  deeper layers.  Coulomb
diffusion might not be negligible in the dense carbon- and He-rich
envelopes of  ultra-massive WDs,  thus possibly preventing  the inward
diffusion of  He toward  the core  and leading  to a  much sharper
transition region than we obtained.
%In this sense, the initial He buffer of our models could not have diffused appreciably to
%deeper layers as we find from our treatment of diffusion. 
We check this  issue by artificially assuming that there  is no inward
chemical diffusion  of He,  finding that  the impact  that Coulomb
separation   could   inflict   on  our   pulsational   inferences   is
moderate.  Finally, all  of  our  WD models  start  to crystallize  at
effective  temperatures   well  above  the  instability   strip,  thus
harboring  a  large  fraction   of  their  stellar  mass  crystallized
(particularly  the  ONe-core WD  for  which  Coulomb interactions  are
larger) when they  reach the ZZ Ceti stage. The  shape of the chemical
profile  is  modified  by  crystallization not  only  in  the  growing
crystallized core  left behind but  also in liquid regions  beyond the
crystallization front. The changes  in the chemical profiles resulting
from  the phase-separation  process  strongly  impact the  theoretical
pulsational  spectrum and  must  be taken  into  account in  realistic
computations  of  the  pulsational  properties  of  ultra-massive  WDs
\citep{2019A&A...621A.100D}.

% For the  He content ,we
%take that given by the evolutionary history of the single progenitor stars, and results 1.8 (1.3) $\times %10^{-4}M_{\odot} $ for the ONe- (CO-) core WD.

%Because of the much larger number of thermal pulses experienced by the %progenitor of the CO WD, this He content ({\bf poner valor})turns out to be %smaller than that characterizing the ONe WD (1.8 $\times 10^{-4}M_{\odot} $) %of the same stellar mass. 

%Several distinctive features are clearly noted, each one formed at different stages in the %progenitor evolution. The innermost CO core is build up during the core He burning and the %following steady He burning during the AGB phase.The evolution along the TP-AGB (in our %simulation, about 1,000 thermal pulses occurred {\bf chequear}) has modeled the outer CO profile. %In particular  the He- and C-rich intershell region is formed during this stage as a result of the
%short-lived He flash convection zone induced by the last He flash we computed. In our %simulation,  the core composition is $\sim 0.45 \%$ $^{12}$C,  and  $\sim 0.55\%$ $^{16}$O,  with %minor traces of mostly  $^{22}$Ne ({\bf chequear los numeros}).  

\begin{figure}
        \centering
        \includegraphics[width=1.\columnwidth]{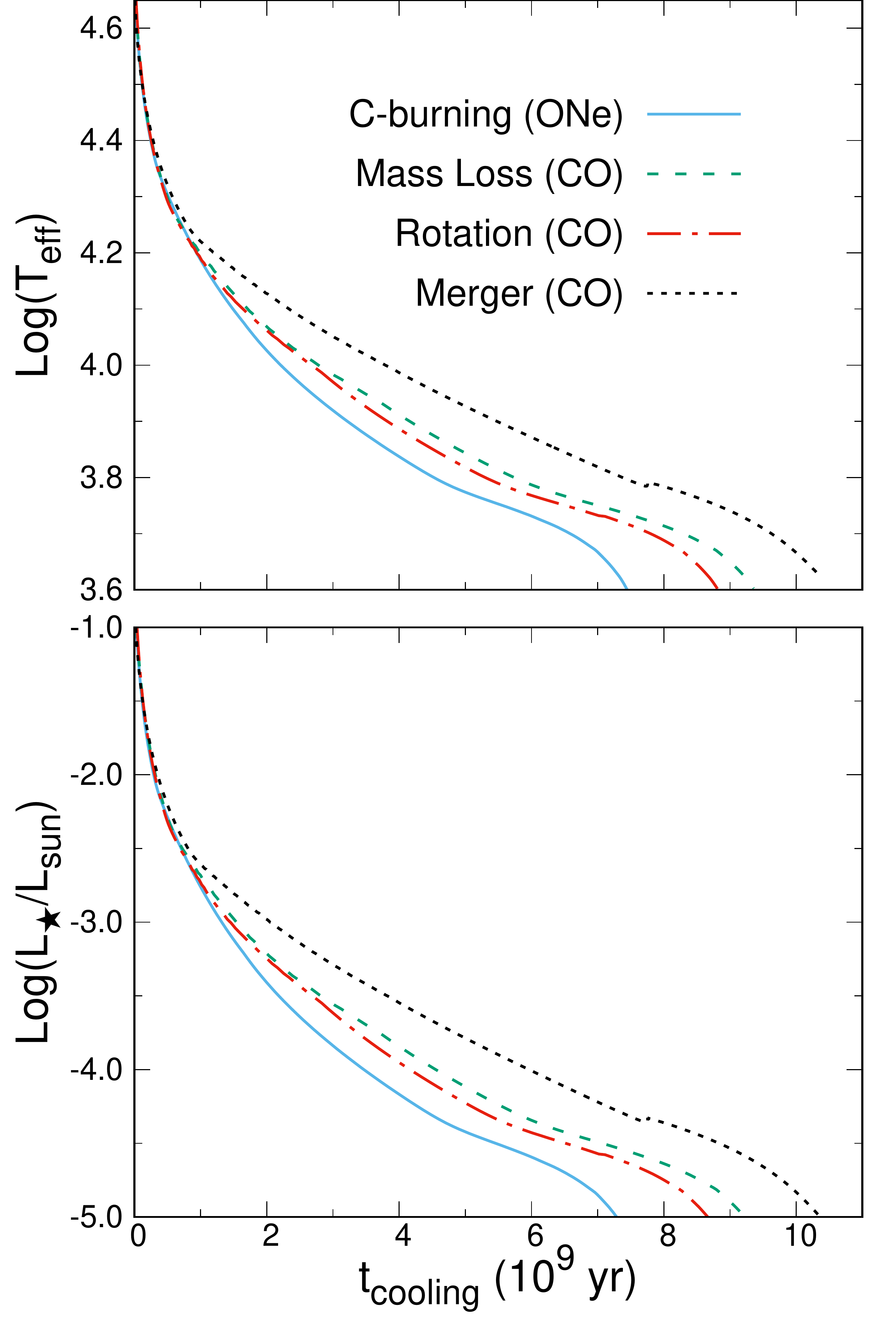}
        \caption{Effective temperature and surface luminosity (upper and bottom panel) in terms of the cooling times for our  $1.156\,M_{\odot}$  WD sequences resulting from the various formation scenarios explored in this paper. Cooling time is counted from the time of WD formation.} 
        \label{DA.eps}
\end{figure}

The evolutionary  scenario through which ultra-massive  WDs are formed
will influence  their cooling  markedly. This  is illustrated  in Fig.
\ref{DA.eps},   which  compares   the  cooling   times  of   our  UMCO
$1.156\,M_{\odot}$ WDs models resulting  from reduced mass-loss rates 
and rotation scenarios, with those of our adopted merged model (dashed, dot dashed,  and doted lines,
respectively) and with those  corresponding to  the ONe-core  WD sequence
resulting  from  carbon  burning during  progenitor  evolution  (solid
line). The cooling  times are set to  zero at the beginning  of the WD
cooling   phase,  when   the  star   reaches  the   maximum  effective
temperature. Gravothermal energy is the  dominant energy source of the
WD. At early stages, say log $L/ L_{\odot} \gtrsim -0.7$ ($\log T_{\rm
  eff} \gtrsim 4.68$), cooling is  dictated by energy lost by neutrino
emission,  which  is  about  the   same  order  of  magnitude  as  the
gravothermal  energy  release. As  WD  cool,  the temperature  of  the
degenerate core decreases and neutrino luminosity drops. At $-$log $L/
L_{\odot}$  =  2.3  -  2.5  ($\log   T_{\rm  eff}$  =  4.23  -  4.28),
crystallization  sets in  at  the  center of  our  UMCO WD  sequences,
resulting in a release of latent  heat and gravitational energy due to
phase separation  of main  core constituents  that impact  the cooling
times. Because  of the larger  coulomb interactions, this  occurs when
the surface luminosity is higher,  log $L/ L_{\odot}=-2$ ($\log T_{\rm
  eff}$ =  4.36), for  the ONe-core WD  sequence, with  the consequent
less relevant impact on the cooling times.
%During  the crystallization  phase  the  surface  luminosity is  larger  than  the
%gravothermal  luminosity.    
Finally, at the  lowest luminosities shown in  Fig.  \ref{DA.eps}, the
temperature   of  the   crystallized  core   drops  below   the  Debye
temperature, and all  of the WD sequences enter  the so-called ``Debye
cooling phase'', characterized by a rapid cooling.

%-2.53   mass loss, -2.35 rotacion, -2.51 merger, -2 One

%To assess the impact of $^{22}$Ne sedimentation we compute an additional sequence in which this processes is fully taken into account, see \cite{2010ApJ...719..612A} and \cite{2016ApJ...823..158C}. The resulting cooling time for the CO-core WD is depicted in Fig.  \ref{DA.eps} with dashed line. For the 
%ONe-core sequence, a much less impact  of $^{22}$Ne sedimentation on the cooling times is expected \footnote{This is because the diffusion coefficient of  $^{22}$Ne  in the liquid core has a strong dependence on the atomic charge $Z$ of the background matter \citep{2010PhRvE..82f6401H} with the result that $^{22}$Ne sedimentation becomes more efficient in a CO-core WD than in a ONe-core WD.} and hence the resulting cooling times are not shown. 

%\begin{figure}
%        \centering
%        \includegraphics[width=1.0\columnwidth]{DB.eps}
%        \caption{Same as Fig. \ref{DA.eps} but for H-deficient, ultra-massive WD sequences.} 
%        \label{DB.eps}
%\end{figure}

\begin{figure}
        \centering
        \includegraphics[width=1.0\columnwidth]{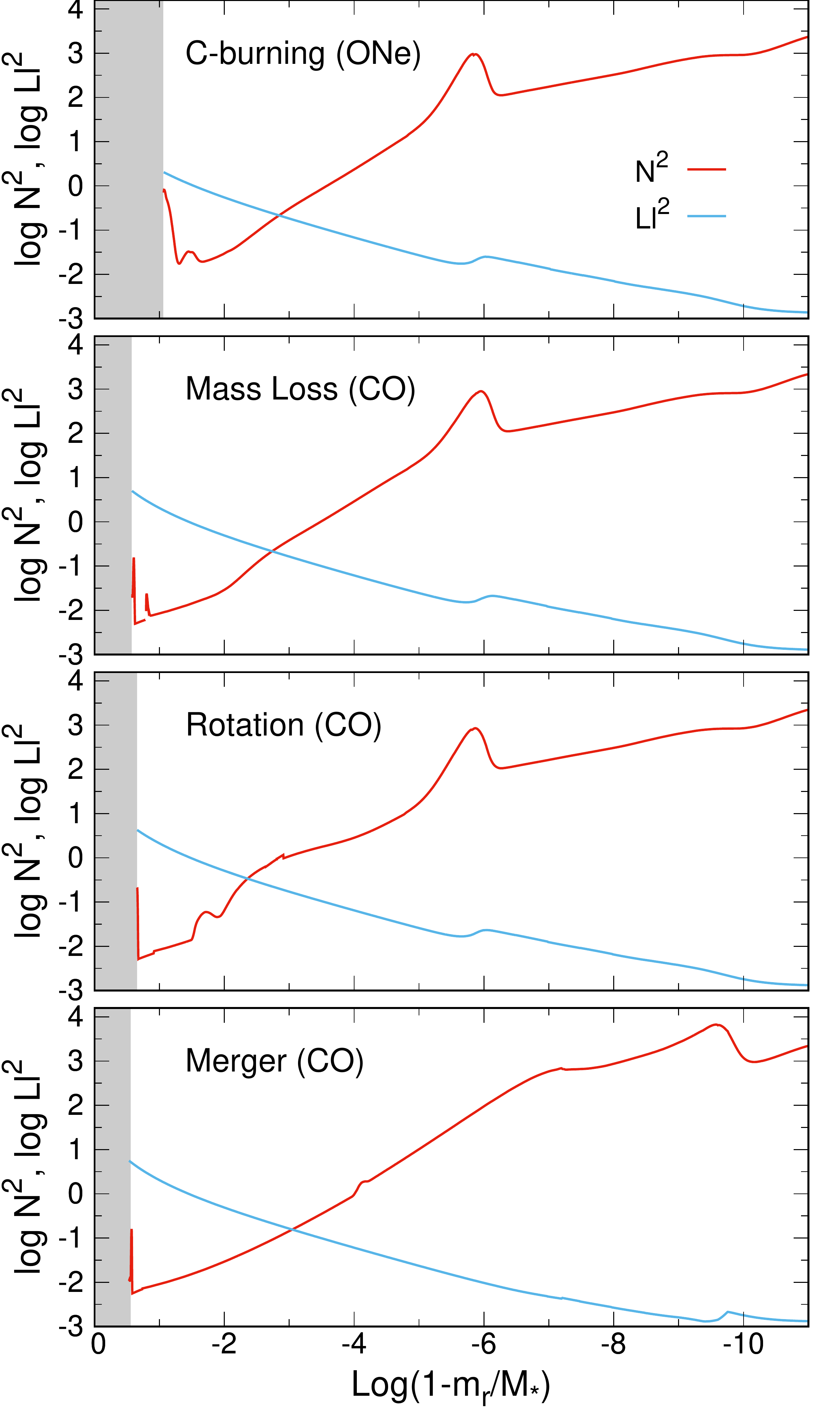}
        \caption{Same as Fig. \ref{perfiles_zz.eps}, but for the
        logarithm of the squared Brunt-Va\"is\"al\"a and Lamb 
frequencies, red and blue lines respectively. The Lamb frequency corresponds to dipole ($\ell= 1$) modes.
In each panel, the gray area corresponds to the crystallized 
part of the models.} 
        \label{bvf}
\end{figure}

Clearly, UMCO WDs evolve slower than  the ONe-core WDs. This is due to
the larger impact  of the energetic of crystallization in  the case of
the CO-core  WDs and  also because  the specific heat  per gram  of CO
composition  is larger  than that  of ONe  composition.  The merged sequence  takes the longest to evolve. The main reason
for this, apart from the fact that the core of these WD models has the
largest  carbon content,  is  that  the He  buffer  is located  at
shallower layers  than in  the other WD  sequences. Hence,  the layers
above the  degenerate core  are more  opaque in  the merger  WD models
(below the WD envelope, carbon opacity is larger than that of He),
thus slowing the cooling rate.  In particular, this sequence takes 1.5
(1.8) times  more to evolve down  to log $L/ L_{\odot}=-3\  (-4)$ than
the ONe-core sequence.
%Finally,  $^{22}$Ne sedimentation yields an extra delay in the cooling times of the CO-core %sequence. In fact, because of its two extra neutrons present,  $^{22}$Ne nucleus slowly
%diffuses in the liquid regions towards the center of
%the WD, releasing sufficient energy to affect markedly the cooling of massive WDs, see \cite{2002ApJ...580.1077D}, \cite{2010ApJ...719..612A}. As shown in  \cite{2010ApJ...719..612A}, $^{22}$Ne sedimentation affects significantly the cooling times of massive WDs coming from progenitors with metallicity above $\approx$ 0.01. We stress that the progenitor metallicity of our ultra-massive CO-core WD sequence is $Z=0.02$. In this case, $^{22}$Ne sedimentation releases enough energy to produce appreciable
%delays, of the order of 15-30 $\%$, in the rate of cooling of the CO-core WD sequence at intermediate luminosities.
%the UMCO WD models from merger needs .... Gyr (.... Gyr)  longer to cool down to log $L/ L_{\odot}=-3$ (-4) than the ONe-core WD models (mejor poner porcentaje). 
Finally,  at log  $L/ L_{\odot}  \sim -4.3$,  there is  a bump  in the
cooling curve  of the merger sequence.  This is a consequence
of the convective mixing of the H envelope with the  underlying
He  buffer (the  mass  fraction of  He in  the
envelope increases up to 80 $\%$). The star has initially an excess of
internal energy  that has to be  radiated to adjust itself  to the new
He-enriched  (more  transparent)  envelope,  with  the  consequent
increase in the cooling time at those stages.

%The impact of the core composition on the cooling times of  H-deficient ultra-massive WDs 
%is illustrated in  Fig. \ref{DB.eps}. The cooling times of the H-deficient WD sequence of  1.16   $M_{\sun}$ with ONe core were taken from \cite{2019A&A...625A..87C}, while the cooling times for the CO-core WD sequences of the same stellar mass were calculated specifically for the purpose of this paper. As in the case of H-rich WDs discussed previously, 
%ultra-massive WDs with CO cores take much longer to evolve than ONe WDs. As discussed in \cite{2019A&A...625A..87C}, at low-luminosities, H-deficient ultra-massive WDs evolve  markedly faster than
%H-rich  WDs.   This is because, at  those stages,  the
%thermal  energy  content of  the  H-deficient  WD  is
%smaller, and   more importantly, because in  these WDs, the  outer layers
%are  more transparent  to radiation.

%\begin{figure}
%        \centering
%%        \includegraphics[width=1.0\columnwidth,angle=-90]{DeltaP-CO-ONe-pilar-12000.ep%s}
%\includegraphics[width=1.0\columnwidth]{fig5.eps}
%        \caption{Forward period spacing ($\Delta \Pi$) in terms of the periods
%  of $\ell= 1$ pulsation $g$ modes for $1.156 M_{\odot}$  WD models at
%  $T_{\rm eff} = 12000$ K with an  CO core (left panel) and a ONe
%  core (right panel). In both models, latent-heat release and chemical
%  redistribution caused by phase separation have  been taken into
%  account during crystallization. The horizontal red-dashed line is
%  the asymptotic period spacing.} 
%        \label{DeltaP-CO-ONe-pilar-12000.eps}
%\end{figure}

\begin{figure}
        \centering
\includegraphics[width=1.0\columnwidth]{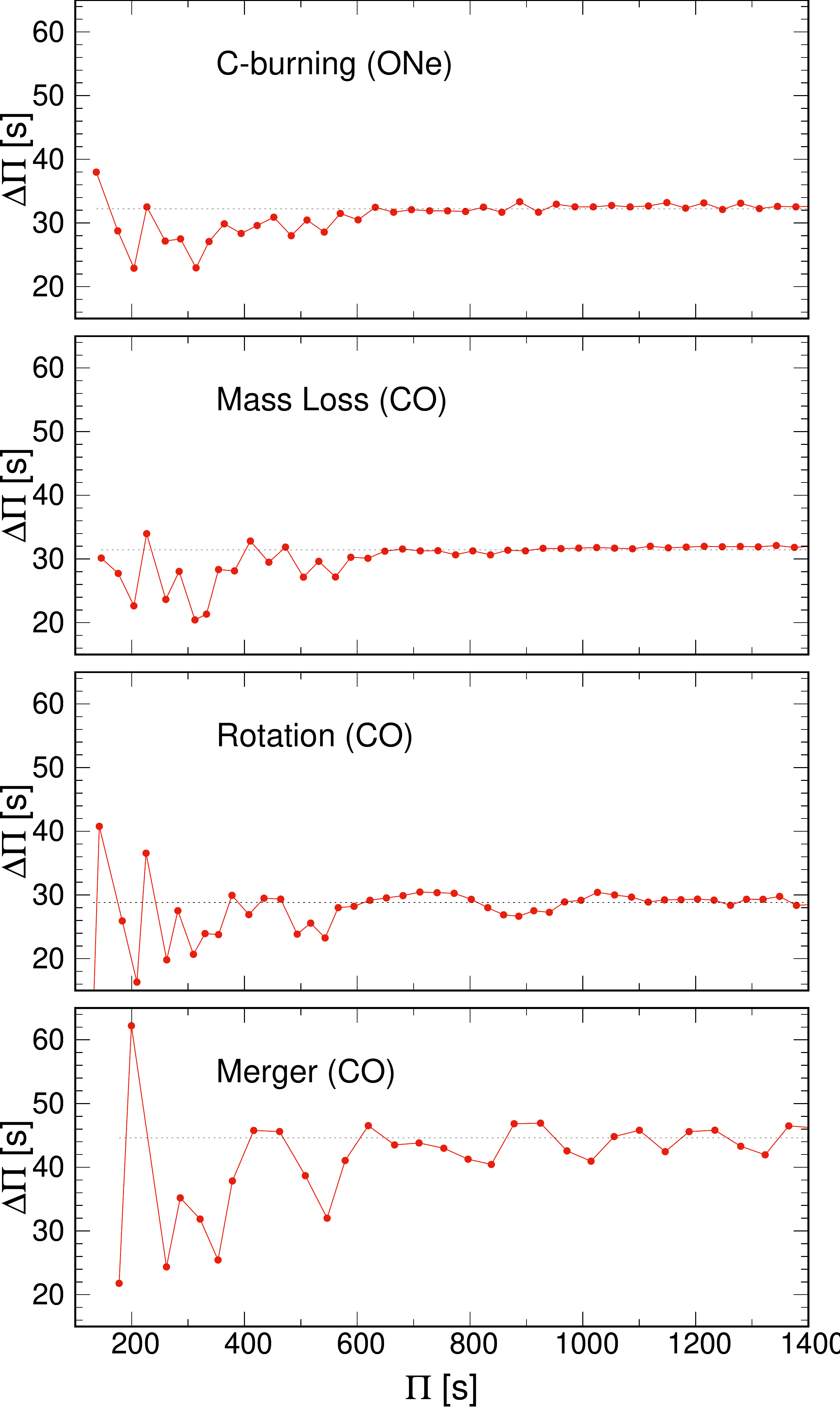}
\caption{Same as Fig. \ref{bvf}, but for the forward period spacing 
        ($\Delta \Pi$) in terms of the periods
  of $\ell= 1$ pulsation $g$ modes. The horizontal black-dashed line is
  the asymptotic period spacing.} 
        \label{DeltaP-CO-ONe}
\end{figure}

%\begin{figure}
%        \centering
%%        \includegraphics[width=1.0\columnwidth,angle=-90]{DeltaP-CO-ONe-pilar-11600.ep%s}
%        \includegraphics[width=1.0\columnwidth]{fig6.eps}        
%        \caption{Same as Fig. \ref{DeltaP-CO-ONe-pilar-12000.eps} but for WD models at % $T_{\rm eff} = 11600$ K .} 
%        \label{DeltaP-CO-ONe-pilar-11600.eps}
%\end{figure}
%
%\begin{figure}
%        \centering
%%        \includegraphics[width=1.0\columnwidth,angle=-90]{DeltaP-CO-ONe-pilar-10000.ep%s}
%        \includegraphics[width=1.0\columnwidth]{fig7.eps}
%        \caption{Same as Fig. \ref{DeltaP-CO-ONe-pilar-12000.eps} but for WD models at % $T_{\rm eff} = 10000$ K.} 
%        \label{DeltaP-CO-ONe-pilar-10000.eps}
%\end{figure}

%Element diffusion smooths the chemical
%profiles, thus shaping the run of the Brunt-V\"ais\"al\"a frequency ---the characteristic frequency of pulsation $g$ modes. 

\section{Pulsation results}
\label{pulsation_results}

Element  diffusion irons  out  any gradient  of chemical  composition,
translating this into  a very smooth shape  of the Brunt-V\"ais\"al\"a
frequency ---the characteristic frequency  of pulsation $g$ modes. The
shape of the Brunt-V\"ais\"al\"a frequency  has a strong impact on the
period  spectrum and  mode-trapping properties  of pulsating  WDs. The
logarithm  of the  squared  Brunt-V\"ais\"al\"a and  Lamb\footnote{The
  Lamb frequency  is the  critical frequency  of pulsation  $p$ modes,
  which  have  yet  to  be  detected  in  pulsating  WDs  until  now.}
frequencies   in  terms   of   the  outer   mass   fraction  for   the
$1.156\,M_{\odot}$ WD  models resulting from the  various evolutionary
scenarios considered  in this paper  are shown in Fig.  \ref{bvf}. The
internal  chemical profiles  of these  models are  those displayed  in
Fig. \ref{perfiles_zz.eps}. In each case, the gray area corresponds to
the crystallized  part of the core.  In this work, we  have considered
that the  eigenfunctions of the  $g$ modes cannot penetrate  the solid
region of the stellar core. For this reason, the propagation region is
restricted to  the fluid  zone, between the  edge of  the crystallized
region  and  the  stellar  surface. This  implies  that  any  chemical
interface located within the crystallized  region in each model has no
relevance  to the  pulsation properties  of  the $g$  modes. In  other
words,  crystallization  prevents   "probing"  the  internal  chemical
structure of  the solid portion of  the core ---left by  the different
formation scenarios of the WD--- through the $g$-mode periods.

In all the  cases, there exists a  dominant feature in the  run of the
Brunt-V\"ais\"al\"a  frequency associated  to  the outermost  chemical
transition region. This chemical  transition region corresponds to the
$^{4}$He/$^{1}$H interface in the case of the WD model coming from the
merger episode, and to the ($^{16}$O/$^{12}$C/$^{4}$He)/$^{1}$H
%$^{16}$O/$^{12}$C/$^{4}$He/$^{1}$H 
interface      for     the      remaining      WD     models      (see
Fig. \ref{perfiles_zz.eps}).  The bump occurs at  a different location
for the case of the WD  model from the merger [$\log (1-M_r/M_{\star})
  \sim  -9.5$] than  for the  other WD  models [$\log(1-M_r/M_{\star})
  \sim -6$]. The large difference in the spatial location of this bump
has  important consequences  for the  mode-trapping properties  of the
models.  We note the absence of any other bump in the Brunt-V\"ais\"al\"a
frequency of the  merger model, in contrast to the  other models. This
implies that  for the  merger model  the mode-trapping  properties are
determined  exclusively by  the only  bump in  the fluid  zone of  the
model, that is,  that due to the  $^{4}$He/$^{1}$H chemical transition
region.

 The mode-trapping properties of WDs are dominated by the presence and
 nature  of  chemical transitions  that are  seen  as bumps  in  the
 Brunt-V\"ais\"al\"a frequency.  Specifically, mode trapping manifests
 itself in that at a given $T_{\rm eff}$ value, the separation between
 consecutive  periods   departs  from   the  mean   (constant)  period
 spacing.  A  tool  for   studying  the  mode-trapping  properties  in
 pulsating  WDs is  the  $\Delta  \Pi -  \Pi$  diagram,  in which  the
 separation  of periods  with consecutive  radial order  $k$ ("forward
 period spacing"  $\Delta \Pi  \equiv \Pi_{k+1}-\Pi_k$) is  plotted in
 terms of the pulsation periods.  In Fig. \ref{DeltaP-CO-ONe}, we show
 the  forward period  spacing in  terms of  the periods  of $\ell=  1$
 pulsation modes for the $1.156\,M_{\odot}$ CO- and ONe-core WD models
 considered in Figs.  \ref{perfiles_zz.eps}  and \ref{bvf}. In all the
 cases, $\Delta  \Pi$ exhibits a regime  of short/intermediate periods
 ($\Pi \lesssim 600$ s) characterized  by maxima and minima typical of
 WD models  harboring one or  more chemical interfaces.   These maxima
 and minima  represent departures  from a constant  period separation,
 which is represented  in the figure by the  asymptotic period spacing
 (horizontal  black-dotted line).  For longer  periods, generally  the
 forward period spacing is much smoother, without strong signs of mode
 trapping, and  quickly converging  to the asymptotic  period spacing,
 except in the  case of the WD model coming  from the merger scenario,
 which exhibits notable maxima and minima even for long periods.
  
A remarkable  difference exists in  $\Delta \Pi$ between the merged  WD  and the  rest of the WD models.  This consists
in that  the amplitude of  trapping (the  magnitude of the  maxima and
minima of $\Delta \Pi$) is notably  larger for this model than for the
remaining   ones.  This   difference  could   be  exploited   in  real
ultra-massive  ZZ Ceti  stars if  it were  possible to  detect several
periods with consecutive  radial orders as to produce  such a diagram,
in order  to distinguish the  origin of the  observed WD, that  is, to
discriminate if the  star has been formed from a  merger episode or is
the result  of single-star evolution.  Another empirical tool  to make
this distinction  could be  elaborated on the  basis that  the average
period spacing for the representative WD  model of a merger episode is
quite larger ($\sim 40 \%$) than  for the template models of the other
scenarios,  as can  be  seen in  Fig.  \ref{DeltaP-CO-ONe}. Indeed,  a
difference  of the  averaged period  spacing of  $\sim 15$  s, as  our
theoretical models  of WDs  with different  origins predict,  could be
easily detected in ultra-massive ZZ  Ceti stars if a sufficient number
of  periods were  detected. This would help  to differentiate  their
possible formation scenarios.

\section{Summary and conclusions}
\label{conclusions}

%We  have  studied  the  evolutionary  and  pulsational  properties  of
%ultra-massive  CO-core  (UMCO) WDs  with  stellar  masses larger  than
%$M_{\rm  WD}  \gtrsim 1.05\,M_\sun  $  resulting  from single  stellar
%evolution and from binary merger.  Predictions have been compared with
%those of the ONe-core WDs  resulting from degenerate carbon burning in
%single-progenitor  stars   with  masses  higher   than  6--9$\,M_\sun$
%\citep[see][for  a   review]{2017PASA...34...56D}.   

We   explore  two
single-star evolution  scenarios for  the formation  of ultra-massive  CO-core  (UMCO) WDs  with  stellar  masses larger  than 
$M_{\rm  WD}  \gtrsim 1.05\,M_\sun$. The
first of these  scenarios involves a reduction in  the mass-loss rates
usually adopted for the evolution of  massive AGB stars.  We find that
in this case, if the minimun CO-core mass for the occurrence of carbon
burning is  not reached  before the  TP-AGB phase, a  UMCO WD  of mass
larger than  $M_{\rm WD}  \gtrsim 1.05\,M_\sun  $ can  be formed  as a
result of  the slow growth  of CO-core  mass during the  TP-AGB phase.
 For this to occur, the mass-loss rates of massive AGB stars need to be reduced by a factor that depends on the poorly known approach to the determination of convection boundaries. Using a search for convective neutrality approach, which favors efficient third dredge-up, imposes the need of mass-loss rates no less  than 5-20 times lower than standard mass-loss rates. On the other hand, when the strict Schwarzschild criterion is used, the required reduction of standard mass-loss rates drops to a factor 2.   We emphasize that the role of  mass loss for the formation of UMCO WDs is conditioned by the efficiency of the third dredge-up. The higher the efficiency of the third dredge-up, the 
 larger the need to decrease mass-loss rates with respect to the standard prescriptions
 . It follows that the calibration of mass loss is heavily dependent on the treatment of mixing.
%For this to occur, the reduction in the mass-loss rates  of massive  AGB stars  need to be  no less  than 5-20 times  lower  than  standard  mass-loss   rates.  
In addition, 
%given  the  lack  of
%observations for  the stellar  mass range in  which are  interested in
%this work, 
this  reduction in the mass-loss rates  cannot be discarded
and is in line with  different pieces of recent observational evidence
that indicate mass-loss rates lower  than expected from current models
\citep[e.g.][]{decin2019}.

The  other single-star  evolution  scenario involves  rotation of  the
degenerate core that results after core He burning at the onset of
the AGB  phase. We have  performed a  series of evolutionary  tests of
solar metallicity models of masses between 4 an 9.5\,$M_{\odot}$, with
core rotation rates $f$ of 0.02,  0.05, 0.1, 0.15, 0.20, and 0.25.  As
shown in Dom96, we find that  the lifting effect of rotation maintains
the maximum temperature at lower values than the temperature necessary
for  off-center carbon  ignition.  In  addition, the  second dredge-up occurs  much
later than in  the case of no  rotation. As a result, the  mass of the
degenerate  CO  core becomes  larger  than  $1.05\,M_\sun$ before  the
TP-AGB. We find that UMCO WDs can  be formed even for values of $f$ as
low as 0.02, and that the range  of initial masses leading to UMCO WDs
widens  as $f$  increases,  whereas  the range  for  the formation  of
ONe-core WDs decreases significantly. Finally we compare our findings
with the predictions from  ultra-massive WDs resulting from the merger of two equal-mass CO-core WDs, by assuming complete mixing between them and a CO core for the merged remnant.

These two single evolution scenarios  produce  UMCO  WDs  with 
different CO profiles and different  He contents.
%To
%isolate the impact on WD evolution of the internal composition predicted by the possible scenarios,  we have adopted the same stellar mass of 1.159\,$M_{\sun}$ for all of our WD sequences. 
Element diffusion  and phase separation of the  CO composition of the core
upon crystallization profoundly alter the initial chemical profiles of
resulting UMCO WDs by the time evolution has proceeded to the domain
of the ZZ ceti stars. 
%We find  that for the UMCO WD resulting from the
%merger, vestiges of inner H  surviving the merger episode float
%to the surface  as a result of gravitational settling,  thus forming a
%thin pure  H envelope and  a pure  He buffer below  it.  
We  find that the evolutionary and pulsational properties of the
UMCO  WDs  formed  through  single evolution  scenarios  are  markedly
different from  those of the ONe-core  WDs resulting from degenerate carbon burning in
single-progenitor  stars \citep[see][for  a   review]{2017PASA...34...56D}
and from those white dwarfs with carbon-oxygen core that
might result from double degenerate mergers. This can eventually be used
to shed  light on  the  core composition  of
ultra-massive WDs  and their  origin. UMCO WDs  evolve markedly  slower than  their ONe
counterparts,\footnote{The evolutionary
sequences are available at \url{http://evolgroup.fcaglp.unlp.edu.ar/TRACKS/tracks.html}}
with cooling times of up  to almost a factor 2 larger.
Stellar populations  with
reliable age inferences could in  principle help to discern the origin
and core composition of ultra-massive WDs.
%Each  scenario of  formation for
%ultra-massive  WDs we  explore  leaves distinctive  signatures in  the
%Brunt-V\"ais\"al\"a frequency and, in turn, in the period spectrum and
%mode-trapping  properties  of pulsating  WDs.  In  particular, the  WD
%models  coming from  the  merger scenario  show  notable maximums  and
%minimums in the forward period spacing for the whole range of relevant
%periods,  in   contrast  with  the   WDs  resulting  from   the  other
%scenarios. 
%We find that both the amplitude of trapping and the average
%of period  spacing are distinctive  features of the UMCO  WD resulting
%from  the  merger,  which  could  potentially  be  exploited  in  real
%ultra-massive ZZ  Ceti stars  if a sufficient  number of  periods were
%detected.  Finally,  another  distinctive feature  that  characterizes
%ultra-massive WDs is their different cooling behavior depending on the
%formation scenario.  UMCO WDs  evolve markedly  slower than  their ONe
%counterparts, with cooling times of up  to almost a factor 2 larger in
%the  case of  UMCO  resulting from  merger\footnote{The evolutionary
%sequences are available at \url{http://evolgroup.fcaglp.unlp.edu.ar/TRACKS/tracks.html}}.  Stellar %populations  with
%reliable age inferences could in  principle help to discern the origin
%and core composition of ultra-massive WDs.

Our investigation strongly suggests the formation  of UMCO WDs from single stellar
evolution. Given recent studies suggesting the formation of ultra-massive
WDs with ONe cores after WD merger \citep{2020arXiv201103546S}, 
and the observational demand 
of the existence of UMCO WDs to explain the large cooling delays of ultra-massive WDs on the Q-branch \citep{2020ApJ...891..160C},
the formation of UMCO via 
single evolution we studied in this paper is of utmost interest. In view  of the  much smaller  cooling rates  of UMCO  WDs with
respect to  ONe-core WDs, this would  have an impact on  the predicted
mass distribution  of massive WDs.  Finally, we note that  even larger
delays in the cooling times of  the UMCO WDs are expected if $^{22}$Ne
sedimentation  is   considered,  see   \cite{2010ApJ...719..612A}  and
\cite{2016ApJ...823..158C}.   This  is   so   because  the   diffusion
coefficient of $^{22}$Ne in the liquid core has a strong dependence on
the    atomic     charge    $Z$     of    the     background    matter
\citep{2010PhRvE..82f6401H},   with    the   result    that   $^{22}$Ne
sedimentation  becomes  more efficient  in  a  CO-core  WD than  in  a
ONe-core WD.

%Considering  that the  merger of  two
%intermediate-mass WDs also  predicts the formation of UMCO  WDs, it is
%not unreasonable to assume that  a large fraction of the ultra-massive
%WD  population could  be  characterized  by CO  cores  instead of  ONe
%ones.  In view  of the  much smaller  cooling rates  of UMCO  WDs with
%respect to  ONe-core WDs, this would  have an impact on  the predicted
%mass distribution  of massive WDs.  Finally, we note that  even larger
%delays in the cooling times of  the UMCO WDs are expected if $^{22}$Ne
%sedimentation  is   considered,  see   \cite{2010ApJ...719..612A}  and
%\cite{2016ApJ...823..158C}.   This  is   so   because  the   diffusion
%coefficient of $^{22}$Ne in the liquid core has a strong dependence on
%the    atomic     charge    $Z$     of    the     background    matter
%\citep{2010PhRvE..82f6401H}   with    the   result    that   $^{22}$Ne
%sedimentation  becomes  more efficient  in  a  CO-core  WD than  in  a
%ONe-core WD.

\begin{acknowledgements}
 
We  wish  to  thank  the  suggestions  and
 comments of an anonymous referee that strongly improved the original
 version of this work. 
Part of  this work was  supported by  PIP
112-200801-00940 grant from CONICET,  by MINECO grants AYA2014-59084-P,
AYA2017-86274-P, and PID2019-109363GB-I00 / AEI / 10.13039/501100011033, by grant G149 from University of La Plata, and by  the AGAUR grant SGR-661/201. 
ARM acknowledges  support  from  the  MINECO under  the  Ram\'on  y  Cajal programme (RYC-2016-20254).
This  research has  made use of  NASA Astrophysics Data System.
\end{acknowledgements}

\bibliographystyle{aa}
\bibliography{ultramassiveCO}

%\newpage

%\begin{appendix}

\end{document}